\documentclass[aps,prr,twocolumn,superscriptaddress]{revtex4-2}
\usepackage{amsmath}
\usepackage{graphicx}
\usepackage[colorlinks=true, linkcolor=blue, citecolor=blue, urlcolor=blue]{hyperref}

\begin{document}


\title{Energy-Efficient Pseudo-Ratchet for Brownian Computers \\through One-Dimensional Quantum Brownian Motion}


\author{Sho Nakade}
\email[]{shonakade.lab@gmail.com}
\affiliation{Advanced ICT Research Institute, National Institute of Information and Communications Technology, Iwaoka 588-2, Kobe 651-2401, Japan}
\affiliation{Department of  Electrical and Electronics Engineering, Mie University, 1577, Kurimamachiya-cho, Tsu, 514-8507, Japan}

\author{Ferdinand Peper}
\email[]{peper@nict.go.jp}
\affiliation{Advanced ICT Research Institute, National Institute of Information and Communications Technology, Iwaoka 588-2, Kobe 651-2401, Japan}

\author{Kazuki Kanki}
\email[]{kanki@omu.ac.jp}
\affiliation{Department of Physics and Nambu Yoichiro Institute of Theoretical and Experimental Physics (NITEP), Osaka Metropolitan University, 3-3-138 Sugimoto, Sumiyoshi-ku, Osaka 558-8585, Japan}

\author{Tomio Petrosky}
\email[]{petrosky@austin.utexas.edu}
\affiliation{Center for Complex Quantum Systems, The University of Texas at Austin, Austin, Texas 78712, USA}
\affiliation{Institute of Industrial Science, The University of Tokyo, 5-1-5 Kashiwa 277-851, Japan}


\date{\today}

\begin{abstract}
Brownian computers utilize thermal fluctuations as a resource for computation and hold promise for achieving ultra-low-energy computations. However, the lack of a statistical direction in Brownian motion necessitates the incorporation of ratchets that facilitate the speeding up and completion of computations in Brownian computers. To make the ratchet mechanism work effectively, an external field is required to overcome thermal fluctuations, which has the drawback of increasing energy consumption. As a remedy for this drawback, we introduce a new approach based on one-dimensional (1D) quantum Brownian motion, which exhibits intrinsic unidirectional transport even in the absence of external forces or asymmetric potential gradients, thereby functioning as an effective pseudo-ratchet.
Specifically, we exploit that quantum resonance effects in 1D systems divide the momentum space of particles into subspaces. These subspaces have no momentum inversion symmetry, resulting in the natural emergence of unidirectional flow. We analyze this pseudo-ratchet mechanism without energy dissipation from an entropic perspective and show that it remains consistent with the second law of thermodynamics.
\end{abstract}


\maketitle


\section{Introduction \label{Introduction}}

In the quest for ultra-low energy computing, there has been growing interest in utilizing thermal fluctuations near equilibrium as computational resources, leading to the concept of ``Brownian computers'' \cite{bennett1982thermodynamics,lee2008brownian,peper2013brownian,lee2016brownian}. Since Brownian motion maintains a momentum distribution in the Maxwell-Boltzmann equilibrium, energy dissipation associated with momentum relaxation can be avoided. Furthermore, because the computation can proceed spontaneously through the particle's thermal diffusion, when combined with logically reversible computation \cite{bennett1973logical,feynman1986quantum,toffoli1980reversible,fredkin1982conservative} that does not involve information erasure, 
it is theoretically possible to approach vanishingly small energy dissipation \cite{landauer1961irreversibility,bennett1982thermodynamics,strasberg2015thermodynamics}.

Within this computational framework, information is physically encoded in the position of a Brownian particle, meaning that the final computational outcome is determined by the particle's arrival at the corresponding designated target position. However, because classical Brownian motion is inherently directionless, it can take an extremely long time for the particle to reach this position \cite{bennett1982thermodynamics,strasberg2015thermodynamics}. More fundamentally, in the absence of external intervention, a Brownian particle cannot deterministically reach the target position as its motion is inherently stochastic, making precise and reliable completion of the computation physically unattainable \cite{norton2013brownian,ray2021non}. This is a well-known consequence of continuous-time Markov processes, in which certain computational operations, such as enforcing a deterministic state transition within a finite time, are fundamentally unimplementable \cite{owen2019number}.

To address these challenges, the Brownian computer paradigm requires external manipulation \cite{peper2013brownian,norton2013brownian,stopnitzky2019physical,ray2021non}.  One approach \cite{lee2008brownian,peper2013brownian,lee2016brownian} involves using a device called \textit{ratchet} that restricts particle motion to a single direction. By transporting particles in a unidirectional manner, the device not only shortens the computation time but also reduces the uncertainty in arrival time at the target position \cite{pal2021thermodynamic}. Consequently, this reduction in arrival time uncertainty enhances the reliability of computation and can mitigate the cost of measurement required for confirming computation completion by reducing the measurement frequency.

However, inducing the directed motion of a Brownian particle requires work to be performed on the particle and thus entails energy dissipation \cite{seifert2012stochastic}. Therefore, both improving computation speed and enhancing reliability of computation involve fundamental trade-offs with minimizing energy consumption \cite{utsumi2022computation,utsumi2023thermodynamic}.

Directional Brownian motion is typically described by an advection-diffusion equation (i.e. the Fokker-Planck equation in the overdamped regime) \cite{risken1996fokker}. The advection term describes the unidirectional motion induced by an external force or a potential, and necessarily involves energy input and dissipation.

The classical Brownian ratchet \cite{magnasco1993forced,astumian1994,astumian1997thermodynamics,reimann2002brownian} is also described by an advection-diffusion equation, where the advection term arises from an asymmetric periodic potential combined with a time-periodic external field. In this framework, the external field does not perform net work on the particle. However, the time-periodic driving maintains the system in a non-equilibrium state, which inherently leads to entropy production and heat dissipation. As a result, the ratchet effect cannot be sustained unless energy is continuously supplied.

In this paper, we propose an approach to overcome these limitations by employing Brownian motion in a one-dimensional (1D) quantum system. The key advantage lies in the one-dimensionality and quantum nature of the system. The particle in our model, remarkably, obeys an advection-diffusion equation even in the absence of any external forces or potential gradients \cite{nakade2020anomalous}:
\begin{align}
  \frac{\partial}{\partial t}
    f^{W}(X,P,&\ t)
  = 
  \nonumber\\
  -\sigma(P)
    \frac{\partial}{\partial X}
  & f^{W}(X,P,t)
  +D(P)
    \frac{\partial^{2}}{\partial X^{2}}
      f^{W}(X,P,t),
  \label{advection-diffusion}
\end{align}
where $f^{W}(X,P,t)$ is the Wigner distribution function of the particle. The first term on the right-hand side represents the advection, describing unidirectional transport with the hydrodynamic sound velocity $\sigma(P)$. The second term corresponds to the diffusion process, spreading the distribution with the diffusion coefficient $D(P)$. Both transport coefficients, $\sigma(P)$ and $D(P)$, depend on momentum.

As a result, this system realizes ratchet-like behavior without conventional energy dissipation mechanisms. Consequently, our 1D quantum system opens up the possibility of mitigating the trade-off between computation efficiency (speed and reliability) and energy dissipation that has long been considered inevitable in classical Brownian computing.

This paper focuses on the analysis of the theoretical physical foundations of Brownian computers, rather than on the specifics of their practical implementation.
In particular, when Brownian particles can be transported in one direction without external manipulation, a question of consistency with the second law of thermodynamics arises, which has been discussed in the context of the Feynman ratchet \cite{feynman1963the} and Maxwell's demon \cite{Szilard:1929aa,sagawa2018second}.
Specifically, we show that it is possible to realize a situation where the advection term overcomes diffusion, leading to the contraction of the spatial distribution of the particle. Since this process occurs spontaneously without energy dissipation, it may appear as negative diffusion. However, we have already proven that the H-theorem holds in this 1D quantum system, ensuring no violation of the second law of thermodynamics \cite{nakade2021anomalous}.

The main objective of this paper is to clarify the specific initial conditions under which the spatial distribution contracts and to analyze the underlying mechanism from the perspective of entropy. By investigating the joint probability distribution of coordinate and momentum, we aim to reveal how quantum effects in a 1D system enable this unique behavior while remaining consistent with the second law of thermodynamics.

Our model considers a quantum particle weakly coupled to 1D lattice vibrations, which form a quantized phonon field acting as a thermal reservoir. Under the stringent spatial confinement of a 1D system, phonon scattering restricts the set of accessible momentum states, partitioning the particle's momentum space into disjoint subspaces \cite{tanaka2009emergence,petrosky2010hofstadter}. Since these subspaces lack momentum inversion symmetry, the equilibrium momentum distribution established within each subspace becomes asymmetric, resulting in a nonzero average velocity and thus unidirectional transport \cite{nakade2020anomalous}.

Crucially, in this 1D setting, dissipation is purely a quantum effect.
If the phonon field is treated in the classical limit, the collision terms vanish, and no dissipation occurs in the one-dimensional classical regime \cite{tanaka2009emergence}. This is because, under the extreme confinement of one dimension, the resonance conditions required for momentum exchange via particle-phonon interactions are severely restricted and cannot be fulfilled classically. In contrast, in a quantum system, the particle can undergo momentum transitions by absorbing or emitting quantized phonons, thereby leading to dissipation.

It should be emphasized that the partitioning of the momentum space, which is a manifestation of the 1D quantum dissipation effect, is independent of temperature. This is because the partitioning stems from the resonance condition in the collision operator and this condition itself is temperature-independent.
This implies that the ratchet-like behavior analyzed in this paper can remain effective over a wide range of temperatures.

Our approach is grounded in microscopic kinetic theory \cite{1977PResiboisMdeLeenery}, enabling us to derive the particle's advection-diffusion equation in the hydrodynamic regime and to analyze its dynamics in detail. This approach is crucial because it allows us to relate transport coefficients, which are often introduced phenomenologically, to underlying microscopic principles.

Our primary finding is that, because of the partitioning of the particle's momentum space into disjoint subspaces, there remains an initial correlation between the particle's coordinate and momentum even after the momentum distribution relaxes to a local equilibrium state. This correlation, quantified by the relative entropy (also known as the Kullback-Leibler divergence \cite{cover1999elements}), ensures that the decrease in spatial distribution entropy does not contradict the second law of thermodynamics. 
Furthermore, the remaining initial correlation between coordinate and momentum implies that the transport direction of the Brownian particle at a given coordinate can be controlled by initial conditions.

This paper is organized as follows. In Section~\ref{1D quantum model and emergence of unidirectional transport without external force}, we introduce a model of a particle coupled to a lattice phonon reservoir, as a prototype for 1D quantum Brownian motion. We focus on the kinetic equation for this system and explain how unidirectional transport emerges. In Section~\ref{Advection-induced spatial contraction in 1D quantum Brownian motion}, we introduce an initial state that leads to the contraction of the spatial distribution and analyze the contraction by examining the time evolution of the mean-square displacement. In Section~\ref{Entropy balance during the spatial distribution contraction}, we explain this spatial contraction scenario from an entropic perspective using relative entropy. Finally, Section~\ref{Concluding remarks} summarizes our results and discusses their potential implications for the practical realization of Brownian computers.

\section{1D quantum model and emergence of unidirectional transport without external force
\label{1D quantum model and emergence of unidirectional transport without external force}}

In this section, we introduce a one-dimensional (1D) quantum Brownian motion model and explain the results of its analysis based on kinetic theory. We demonstrate that the partitioning of momentum space into subspaces—the fundamental origin of all the peculiar phenomena discussed in this paper—arises due to the resonance condition of the collision operator in the one-dimensional quantum system.

This section primarily serves as a review of earlier work; thus, only key results are provided. For a more detailed analysis, the reader is referred to our previous publications \cite{tanaka2009emergence,tay2011band, nakade2020anomalous}. In what follows, we explain from a kinetic perspective why 1D quantum Brownian motion exhibits a unidirectional transport behavior. Readers who are familiar with the underlying mechanisms may skip directly to Sec.~\ref{Advection-induced spatial contraction in 1D quantum Brownian motion}.

In kinetic theory, transport coefficients such as sound velocity and diffusion coefficient, which characterize macroscopic transport properties, are defined in the hydrodynamic regime \cite{prigogine1962non, 1977PResiboisMdeLeenery}. This regime refers to situations where the characteristic spatial scale of particle distribution variations is much larger than the mean free path of the particle. 
In this regime, microscopic collisions occur frequently enough for the momentum distribution to equilibrate locally, while changes in the spatial distribution occur over a much larger length and time scale, governed by macroscopic dynamics.

In this regime, the irreversible collision term dominates the kinetic equation, while the reversible flow term acts as a small perturbation. Consequently, a perturbative analysis of the eigenvalue problem of the collision operator becomes an effective approach. After the rapid relaxation modes have decayed, the remaining hydrodynamic transport modes (e.g., hydrodynamic sound wave and diffusion) emerge through the first- and second-order perturbations applied to the zero-eigenstate of the collision operator. This framework thereby connects the macroscopic transport coefficients to the underlying microscopic dynamics.

In a classical gas system \cite{1977PResiboisMdeLeenery}, a nonzero sound velocity arises due to the five-fold degeneracy of the zero-eigenstates corresponding to the five collisional invariants—particle number, momentum (in three spatial dimensions), and energy. The flow term lifts the degeneracy of these five zero-eigenstates, transforming them into a new basis represented as linear combinations of these states. In this process, the collisional invariants associated with momentum play a crucial role. Their corresponding zero-eigenstates are odd functions in momentum space, and this odd-parity symmetry underpins the emergence of a nonzero sound velocity.

In contrast, classical Brownian motion has only one collisional invariant, the particle number, and its zero-eigenstate corresponds to a symmetric Maxwellian distribution. 
Because of this symmetry, coupling with the flow term results in zero sound velocity and thus no sound mode.

However, in the 1D quantum system, a nonzero sound velocity can emerge even though particle number is the only collisional invariant. This phenomenon occurs because the resonance condition in 1D quantum system partitions the momentum space into multiple subspaces \cite{tanaka2009emergence,petrosky2010hofstadter}. Each subspace has a nondegenerate zero-eigenstate. Since these subspaces do not exhibit momentum-inversion symmetry, the zero-eigenstate within each subspace yields an asymmetric Maxwellian distribution for momentum inversion \cite{nakade2020anomalous}. Consequently, the sound velocity defined within each subspace attains a nonzero value.

\subsection{Model and Kinetic equation
\label{Model and Kinetic equation}}

The model under consideration is a 1D quantum particle coupled to phonons. 
Despite possessing both translational and momentum-inversion symmetries, this system exhibits unidirectional transport of the particle without external forces \cite{tanaka2009emergence}.

Our Hamiltonian is explicitly given by:
\begin{align}
  H&=H_{0}+gV,
  \label{Total Hamiltonian}\\
  H_{0} & =
  \sum_{p}
    \varepsilon_{p}
    |p\rangle\langle p|
  +
  \sum_{q}
    \hbar\omega_{q} a_{q}^{\dagger}a_{q},
    \label{H_0}\\
  gV & =
  \sqrt{\frac{2\pi}{L}}
  \sum_{p,q}
    gV_{q}|p+\hbar q\rangle\langle p|
    (a_{q}+a_{-q}^{\dagger}).
    \label{V}
\end{align}
Here, $|p\rangle$ represents the momentum state of the particle, orthonormalized by the Kronecker delta, while $\hat{a}_{q}^{\dagger}$ and $\hat{a}_{q}$ are phonon creation and annihilation operators with wave number $q$. 
The dimensionless coupling constant $g$ is a parameter used to track the order of the perturbation expansion. After finishing the weak-coupling approximation, we set $g = 1$. The parameter $L$ represents the length of the system.

The dispersion relations and interaction potential take the following forms:
\begin{equation}
  \varepsilon_{p}=\frac{p^{2}}{2m},\ \ \omega_{q}=c|q|,
  \label{dispersion relation}
\end{equation}
\begin{equation}
  gV_{q}=
    g\Delta_{0} |q| 
    \sqrt{
      \frac{\hbar}{4\pi\rho_{M}\omega_{q}}
         },
  \label{deformation potential}
\end{equation}
where $m$ is the effective mass of the particle, $c$ is the propagation speed of acoustic phonons in the 1D chain, $g\Delta_{0}$ is the coupling constant, and $\rho_{M}$ is the mass density of the system.

This model is commonly known as the Davydov Hamiltonian \cite{davydov1982solitons, scott1992davydov,tanaka2009emergence} in the context of biological systems, where it describes the propagation of a vibrational exciton weakly coupled to acoustic phonons of underlying lattice along a one-dimensional $\alpha$-helical protein molecular chain.
Additionally, this model and the same interaction mechanism~\eqref{deformation potential}, known as the deformation-potential interaction, are widely used in semiconductor physics to describe the coupling between a free electron and acoustic phonons \cite{mahan1993many}.

For consistency, all figures in this paper are presented in a dimensionless form, using a unit system where the coordinate and momentum units are defined as $x_{\rm u}=\hbar/mc$, $p_{\rm u}=mc$, and the time unit is $t_{\rm u}=\hbar/mc^2$. The temperature is expressed in units of $T_{\rm u}=mc^2/k_{\rm B}$. Here, $k_{\rm B}$ is the Boltzmann constant. This choice of units corresponds to setting $m=1$, $c=1$, $\hbar=1$, and $k_{\rm B}=1$.

In this system, we define the characteristic scale of the diffusion coefficient as
\begin{equation}
   D^{\ast}=\frac{\hbar^2\rho_M c^3}{ m \Delta_0^{2}},
   \label{D_u}
\end{equation}
which is determined from the microscopic kinetic analysis of this model \cite{nakade2020anomalous}.
To simplify the numerical analysis and highlight the intrinsic dynamics of the system, we choose the parameters $\rho_M$ and $\Delta_0$ such that $D^{\ast}$ becomes equal to the unit of diffusion $D_{\rm u}=x_{\rm u}^2/t_{\rm u}=\hbar/m$, i.e., $D^{\ast}/D_{\rm u}=1$.

Due to the periodic boundary condition imposed with a period $L$,
the discrete momenta and wave numbers are given by $p/\hbar = 2\pi j / L$ and $q = 2\pi j^{\prime} / L$, respectively,  where $j, j^{\prime} = 0, \pm1, \pm2, \dots$. 
We are concerned with the thermodynamic limit ($L \rightarrow \infty$), and therefore we replace the summation over a momentum or wave number with an appropriate integration.

The time evolution of the total system is governed by  the Liouville-von Neumann equation (quantum Liouville equation),
\begin{equation}
  i\frac{\partial}{\partial t}
    \rho(t)
  ={\cal L}_{H}
    \rho(t),
\end{equation}
where  $\rho(t)$ is the density operator of the total system, and the Liouville-von Neumann operator (Liouvillian) ${\cal L}_{H}$ is defined as a commutation relation with Hamiltonian $H$ by
${\cal L}_{H}\rho(t)=[H,\rho(t)]/\hbar$.
We consider the time evolution of the reduced density operator for the particle,
\begin{equation}
  f(t)
  :=
  {\rm Tr}_{{\rm ph}}
    [\rho(t)],
\end{equation}
where ${\rm Tr}_{{\rm ph}}$ represents the trace taken over all phonon modes. 
We assume that the phonons are in thermal equilibrium at temperature $T$.

We introduce the Wigner basis \cite{prigogine1962non,petrosky1997liouville} as the representation basis for the reduced density operator $f(t)$:
\begin{equation}
  f_{k}(P,t)
  :=
    \left(P+\frac{\hbar k}{2}\right|
     f(t)
    \left|P-\frac{\hbar k}{2}\right).
  \label{Wigner representation}
\end{equation}
Here, the round bracket is defined by $|p):=\sqrt{L/2\pi\hbar} |p\rangle$,
and normalized by the Dirac delta function in the limit $L\rightarrow\infty$ as
\begin{equation}
  (p|p^{\prime})=\delta(p-p^{\prime}).
  \label{delta normalize}
\end{equation}
The Fourier transform of $f_{k}(P,t)$ with respect to $k$ defines the Wigner distribution function in ``phase space'':
\begin{equation}
  f^{W}(X,P,t)
  :=
  \frac{1}{2\pi}
    \int_{-\infty}^{\infty}  dk\ 
      e^{ikX}
      f_{k}(P,t).
        \label{Wigner function}
\end{equation}
Note that the $k=0$ component, $f_{0}(P,t)$, corresponds to the momentum distribution, while the $k\neq0$ components represent spatial inhomogeneities.

In the weak coupling approximation, $f_{k}(P,t)$ satisfies a Markovian kinetic equation \cite{prigogine1962non,petrosky1996poincare,petrosky1997liouville,tanaka2009emergence},
\begin{equation}
  i\frac{\partial}{\partial t}
    f_{k}(P,t)
  =
  {\cal K}_P^{(k)}
    f_{k}(P,t),
  \label{kinetic equation}
\end{equation}
where ${\cal K}_P^{(k)}$ is the second-order approximation of the collision operator with respect to the particle-phonon coupling. It is expressed as \cite{tanaka2009emergence}
\begin{equation}
   {\cal K}_P^{(k)}= k\frac{P}{m}+{\cal K}_P^{(0)},
   \label{collision op}
\end{equation}
where the flow term $kP/m$ represents the time-reversible dynamics, and the collision term ${\cal K}_P^{(0)}$ governs the irreversible relaxation of the momentum distribution.

\begin{widetext}
The explicit form of ${\cal K}_P^{(0)}$ is given by:
\begin{align}
  {\cal K}_P^{(0)}
 & =-g^{2}
      \frac{2\pi i}{\hbar^{2}}
        \int\!\!dq
          |V_{q}|^{2}
            \left\{
              \delta
                \left(
                  \frac{
                        \varepsilon_{P}-\varepsilon_{P+\hbar q}
                       }{\hbar}
                       +\omega_{q}
                \right)
                n(q)
             +\delta
               \left(
                 \frac{
                      \varepsilon_{P-\hbar q}-\varepsilon_{P}
                      }{\hbar}
                      +\omega_{q}
               \right)
                 [n(q)+1]
            \right\} 
            \nonumber \\
   & \ \ \ 
  +g^{2}
    \frac{2\pi i}{\hbar^{2}}
      \int\!\!dq
        |V_{q}|^{2}
          \left\{ 
            \delta
              \left(
                \frac{
                      \varepsilon_{P-\hbar q}-\varepsilon_{P}
                      }{\hbar}
                +\omega_{q}
              \right)
                n(q)
                  \exp
                     \left[
                            -\hbar q\frac{\partial}{\partial P}
                     \right]
  \right.\nonumber \\
 & \ \ \ \ \ \ \ \ \ \ \ \ \ \ \ \ \ \ \ \ \ \ \ \ \ \ \ \ \ \ \ \ \ \ 
  \left.
  +\delta
    \left(
      \frac{
           \varepsilon_{P}-\varepsilon_{P+\hbar q}
           }{\hbar}
      +\omega_{q}
    \right)
      [n(q)+1]
        \exp
                     \left[
                            \hbar q\frac{\partial}{\partial P}
                     \right]
          \right\},
          \label{concrete form of collision operator}
\end{align}
\end{widetext}
where $n(q)$ is the average number of phonons with wave number $q$, which follows the Bose-Einstein distribution, given by
\begin{equation}
  n(q):=
    \frac{1}{\exp[\hbar\omega_{q}/k_{\rm B}T]-1}.
    \label{n(q)}
\end{equation}

Reflecting the quantum nature of the system, the collision term~\eqref{concrete form of collision operator} acts as a momentum-shift operator, namely, $\exp(\pm\hbar q\cdot\partial/\partial P)f(P)=f(P\pm\hbar q)$.
An important point to note is that in the classical limit for the phonon field, where Planck's constant $\hbar$ is negligible compared to the action variable $J_q$ of the phonon modes  (i.e., $\hbar \ll J_q$), the collision term (\ref{concrete form of collision operator}) takes the form $\int dq\ q\delta(q)$ and vanishes  \cite{tanaka2009emergence}. 
This highlights that the dissipation in this one-dimensional weak coupling system is a purely quantum effect, arising from the quantum nature of the phonon field.

As shown in Ref.~\cite{tanaka2009emergence}, the eigenvalues of the collision operator~\eqref{collision op} completely coincide with the eigenvalues of the original Liouvillian ${\cal L}_H$ in a non-Hilbert space.
Therefore, the information obtained from the eigenvalue problem of the collision operator of this kinetic equation offers fundamental information about the microscopic properties possessed by the original dynamical system based on the principles of physics, and is by no means a phenomenological property obtained by artificial manipulation based on phenomenological arguments such as coarse-graining.

\subsection{Partitioning of momentum space by resonance condition in 1D quantum system 
\label{Partitioning of momentum space by resonance condition in 1D quantum system }}

Let us now focus on the resonance condition represented by the delta-function within the collision term~\eqref{concrete form of collision operator} to discuss the characteristic momentum transitions in this 1D quantum system.

If the momentum after a phonon absorption or emission is written as $P^{\prime}=P\pm\hbar q$, the resonance condition between the two momenta $P$ and $P^{\prime}$ can be expressed as
\begin{align}
                0= &\varepsilon_{P^{\prime}}-\varepsilon_{P}
                      \pm{\hbar}\omega_{|P^{\prime}-P|/\hbar}
                      \nonumber\\
                  = &\frac{1}{2m}(P^{\prime}-P)(P^{\prime}+P\pm 2mc\cdot\mathrm{sgn}(P^{\prime}-P)),
        \label{resonance condition}
\end{align}
where $\mathrm{sgn}()$ is a sign function, defined as $\mathrm{sgn}(x) = +1$ if $x > 0$, $-1$ if $x < 0$, and $0$ if $x = 0$.
This equation indicates that an exciton in the momentum state $P$ transitions to only two possible momentum states, $P^{\prime}=-P\pm2mc$, through the absorption or emission of a phonon.
Starting with momentum $P_0$, all the momenta successively connected by the collision term are enumerated in the form of a recurrence relation as
\begin{equation}
  P_{\nu\pm1}
  =
  -P_{\nu} \pm(-1)^{\nu} 2mc, 
\end{equation}
where $\nu$ is any integer.
The solution of this recursive formula yields the following subset of discrete momenta with an initial value $P_0$:
\begin{equation}
  P_{\nu}=(-1)^{\nu}(P_{0}-2\nu mc).
  \ \ \ \ 
  (\nu=0,\pm1,\pm2,\cdots)
  \label{momentum subset}
\end{equation}

If we choose an initial momentum $P_0$, the set of subsequent momenta is uniquely determined, forming a distinct subspace. However, different choices of $P_0$ may produce overlapping subsets. 
By restricting $P_0$ to a certain range, 
\begin{equation}
  -mc\leq P_{0}\leq mc,
  \label{P_0 range}
\end{equation}
we can avoid such overlaps and ensure that each subspace is uniquely associated with a specific $P_0$.
From here on, we denote the subspace associated with the representative momentum $P_0$ as ${\mathcal S}_{P_0}$. 
When necessary, we write $P_{\nu}(P_0)$ to emphasize its dependence on $P_0$.

A significant feature of the subspace ${\mathcal S}_{P_0}$ is that, except for ${\mathcal S}_{P_0=0}$, they do not exhibit momentum inversion symmetry. 
Specifically, applying the momentum inversion operation $P \mapsto -P$ to a discrete momentum in the subspace ${\mathcal S}_{P_0}$ results in a momentum not contained within the same subspace:
\begin{equation}
     P_\nu(P_0)
     \xrightarrow{P \mapsto -P}-P_\nu(P_0)\notin {\mathcal S}_{P_0},
     \ \ \ \ \ (\forall P_0\neq0)
     \label{momentum symmetry breaking}
\end{equation}
Instead, the inverted momentum belongs to the $-\nu$-th discrete momentum in the subspace ${\mathcal S}_{-P_0}$.
\begin{align}
    - P_\nu(P_0)
    &=P_{-\nu}(-P_0)\in {\mathcal S}_{-P_0}
    \neq {\mathcal S}_{P_0}.
     \ \ \ \ \ (\forall P_0\neq0)
        \label{a paired subspace}
\end{align}
The existence of a mirror subspace ${\mathcal S}_{-P_0}$ paired with ${\mathcal S}_{P_0}$ ensures that momentum inversion symmetry is preserved through the entire momentum space.

In the 1D quantum system, momentum space is partitioned into disjoint subspaces, and the particle's momentum relaxation is confined within each subspace, independent of the others.
Consequently, the equilibrium distribution corresponding to a zero-eigenfunction of the collision term~\eqref{concrete form of collision operator} (i.e., a collisional invariant representing particle number) is given as
\begin{align}
  \varphi_{P_{0}}^{{\rm eq}}(P_{\nu})
  =\frac{
        \exp[-\varepsilon_{_{P_{\nu}(P_0)}}/k_{{\rm B}}T]
       }
       {
        \sum_{\mu=-\infty}^{\infty}
          \exp[
               -\varepsilon_{_{P_{\mu}(P_0)}}/k_{{\rm B}}T
              ]
       }.
  \label{zero eigenstate}
\end{align}
Due to the asymmetry~\eqref{momentum symmetry breaking}, this distribution is an asymmetric Maxwellian distribution for momentum inversion.

In contrast, in systems with two or more dimensions, the freedom of the collision angle makes the number of momentum transitions that satisfy the resonance condition infinite. 
As a consequence, momentum space is not partitioned into distinct subspaces.

Moreover, because the resonance condition~\eqref{resonance condition} is temperature-independent, this characteristic feature of momentum transitions persists under all thermal conditions.

\subsection{Transport properties in the hydrodynamic regime}

We summarize the transport properties of the 1D quantum Brownian particle in the hydrodynamic regime, where unidirectional transport emerges due to the lack of momentum inversion symmetry within each subspace.

In kinetic theory \cite{prigogine1962non, 1977PResiboisMdeLeenery},
macroscopic hydrodynamic transport modes are analyzed microscopically by examining the eigenvalue problem of the collision operator ${\cal K}_P^{(k)}$ in the kinetic equation~\eqref{kinetic equation}, with the flow term $kP/m$ treated as a small perturbation to the primary collision term ${\cal K}_P^{(0)}$.
The two transport coefficients, i.e., the hydrodynamic sound velocity and diffusion coefficient, are defined respectively as the coefficients of the first- and second-order terms of the perturbation expansion with respect to the wave number $k$ of the eigenvalue of ${\cal K}_P^{(k)}$ around the unperturbed zero-eigenvalue.

Reflecting the characteristic of the 1D quantum system that momentum relaxation occurs within each subspace, different transport modes exist for each subspace. 
Specifically, in the hydrodynamic regime, after the momentum relaxation time $\tau_{\mathrm{rel}}$, the Fourier component of Wigner function with a momentum $P_\nu$ belonging to a subspace ${\mathcal S}_{P_0}$ behaves \cite{nakade2020anomalous} as follows:
\begin{align}
  &f_{k}(P_{\nu},t\gtrsim \tau_{{\mathrm{rel}}})
  \simeq
  \nonumber\\
  &
    e^{-i k\sigma(P_0)t}
    e^{-k^2 D(P_0)t}     
     \varphi_{P_{0}}^{{\rm eq}}(P_{\nu})
      \sum_{\mu=-\infty}^{\infty}
        f_{k}(P_{\mu}(P_0),0).
  \label{fk(t) at local equilibrium}
\end{align}
Here, $\sigma(P_0)$ and $D(P_0)$ represent the hydrodynamic sound velocity and diffusion coefficient, respectively, within the subspace ${\mathcal S}_{P_0}$.
This formulation highlights the dependence of transport dynamics on the initial state $f_{k}(P_{\mu}(P_0),0)$, reflecting the subspace-confined nature of the relaxation process.

The expression~\eqref{fk(t) at local equilibrium} is valid under local equilibrium, where the momentum distribution within each subspace has already relaxed to the Maxwellian as defined by Eq.~\eqref{zero eigenstate}.
During this stage, spatial relaxation is still in progress, governed by the small perturbation (approximated up to the second order in $k$) to the zero-eigenvalue of the collision term. 
The first-order perturbation, corresponding to the real part of the eigenvalue of ${\cal K}_P^{(k)}$, 
introduces a reversible wave-like phase factor (i.e., the hydrodynamic sound propagation).
In contrast, the second-order perturbation, corresponding to the imaginary part, 
provides an irreversible exponential decay (i.e., the diffusion).

The hydrodynamic sound velocity $\sigma(P_0)$ takes the form of the average velocity of the particle in the momentum equilibrium state within the subspace ${\mathcal S}_{P_0}$:
\begin{align}
  \sigma(P_{0})
  =\!\!
   \sum_{\nu=-\infty}^{\infty}
     \frac{P_{\nu}}{m}
       \varphi_{P_{0}}^{{\rm eq}}(P_{\nu})
  =
   \frac
   {
    \sum_{\nu=-\infty}^{\infty}
        \frac{P_{\nu}}{m}
        \exp \left(
                    -\frac{P_{\nu}^2}{2mk_{\rm B}T}
             \right)
   }
   {
    \sum_{\mu=-\infty}^{\infty}
        \exp \left(
                    -\frac{P_{\mu}^2}{2mk_{\rm B}T}
             \right)
   }.
       \label{sound velocity 2}
\end{align}
The lack of momentum inversion symmetry within the subspace, as shown in Eq.~\eqref{momentum symmetry breaking}, prevents the cancellation of $\nu$ and $-\nu$ components in the summation. Consequently, $\sigma(P_0)$ takes non-zero positive or negative values, except for the symmetric case $P_0 = 0$, where $\sigma(P_0) = 0$.

The sound velocity $\sigma(P_0)$ gradually vanishes as the temperature increases, implying that the particle's motion approaches classical behavior. This is because, at higher temperatures, a greater number of discrete momentum states $P_{\nu}$ become thermally excited, and the positive and negative contributions in Eq.~\eqref{sound velocity 2} tend to cancel each other out. 
In Appendix~\ref{appendix phonon quantum effect}, we further analyze the high-temperature condition under which the sound velocity asymptotically vanishes, relating it to the crossover in phonon statistics from quantum to classical. The analysis yields the following condition under which the sound velocity asymptotically vanishes (see Eq.~\eqref{high-temperature condition for the particle} in Appendix~\ref{appendix phonon quantum effect}): 
\begin{equation}
   T\gg T_{\rm u},
   \label{crossover temperature}
\end{equation}
where $T_{\rm u}=mc^2/k_{\rm B}$ is the unit temperature used throughout this paper.

The infinite series summations in Eq.~\eqref{sound velocity 2} can be expressed in terms of elliptic theta functions as follows:
\begin{align}
&\sigma(P_{0},T)
   =
   \nonumber\\
    &
    -\frac{k_{{\rm B}}T}{8mc}
       \frac{
               \left.
               \frac{\partial}{\partial p}
                       \vartheta_{3}\bigl(
                                                    \pi p,\ 
                                                    \tilde{q}_{\scalebox{0.5}{$T$}}
                                            \bigr)
               \right |_{p=\frac{{\tilde P}_0 }{4}}
                                            \ 
              -
              \left.
              \frac{\partial}{\partial p}
                       \vartheta_{3}\bigl(
                                                   \pi p,\   
                                                   \tilde{q}_{\scalebox{0.5}{$T$}}
                                            \bigr)
              \right |_{p=\frac{1}{2} (\frac{{\tilde P}_0 }{2}-1)}
              }
             {
              \vartheta_{3}\left(\frac{\pi{\tilde P}_0 }{2},\ 
                                         \tilde{q}^4_{\scalebox{0.5}{$T$}}
                                  \right)
               },
\label{eq: the sound velocity expressed by the theta function}
\end{align}
where $\vartheta_{3}(\tilde{z},\tilde{q})$ represents an elliptic theta function, defined as
\begin{equation}
  \vartheta_{3}(\tilde{z},\tilde{q})
  :=
  1+2\sum_{n=1}^{\infty}\tilde{q}^{n^2}\cos(2n\tilde{z}),
  \label{eq:definition of the elliptic theta function}
\end{equation}
with the notations ${\tilde P}_0$ and $\tilde{q}_{\scalebox{0.5}{$T$}}$ being defined as
\begin{equation}
  {\tilde P}_0 := \frac{P_0}{mc}, \ \ \ \ \
   \tilde{q}_{\scalebox{0.5}{$T$}} := \exp \left[  -\frac{ \pi^2 }{8}
   \frac{k_{\rm B} T}{ mc^2} \right].
\end{equation}
This expression facilitates a more detailed analysis of the hydrodynamic sound velocity behavior (for details, see \cite{nakade2020anomalous}).

On the other hand, the diffusion coefficient $D(P_{0})$ always takes positive values,
\begin{equation}
  D(P_{0})>0,
  \label{D(P_0)>0}
\end{equation}
and increases monotonically with temperature.
Since these coefficients depend only on the subspace itself and not on the specific choice of the initial momentum within the subspace, the following relations hold:
\begin{equation}
  \sigma(P_0)
  =
  \sigma(P_{\nu}),
  \ \ \ \ 
  D(P_0)
  =
  D(P_{\nu}).
  \ \ \ 
  (\nu=0,\pm1,\pm2,\cdots)
  \label{momentum dependency of the transport coefficients}
\end{equation}
For detailed analyses of the transport coefficients, see Ref. \cite{nakade2020anomalous}.

The Wigner distribution function~\eqref{Wigner function} of the reduced density operator~\eqref{fk(t) at local equilibrium} is obtained as
\begin{equation}
  f^{W}(X,P_{\nu},t\gtrsim\tau_{{\mathrm{rel}}})
  \simeq
  \chi_{P_{0}}(X,t)
    \varphi_{P_{0}}^{{\rm eq}}(P_{\nu}),
\label{Wigner distribution at local eq}
\end{equation}
where
\begin{align}
  &\chi_{P_{0}}(X,t) 
  :=
  \nonumber \\
  &\frac{1}{2\pi}
    \int_{-\infty}^{\infty}\!dk\ 
      e^{
         ik\{X-\sigma(P_{0})t\}
         }
      e^{
         -k^{2}D(P_{0})t
         }
      \sum_{\mu=-\infty}^{\infty}
        f_{k}(P_{\mu}(P_0),0).
   \label{chi at local eq}
\end{align}
Note that because $\chi_{P_{0}}(X,t)$ depends on $P_0$, the Wigner distribution function cannot be separated into functions of $X$ and of $P$, even after momentum relaxation.
Although Eq.~\eqref{Wigner distribution at local eq} is defined on the discrete momenta $P_{\nu}$, by varying $P_0$ within the range of~\eqref{P_0 range}, the corresponding $P_{\nu}$ can take any real value. Thus, the function~\eqref{Wigner distribution at local eq}, together with Eq.~\eqref{momentum dependency of the transport coefficients} can be defined over the continuous momentum space $P$.

From here on, we treat the Wigner distribution function as defined on the continuous momentum $P$.
Integrating over the continuous momentum space $P$ is equivalent to summing over all discrete momenta $P_{\nu}$ within subspace ${\mathcal S}_{P_0}$, and then integrating over $P_0$.
\begin{align}
	\int_{-\infty}^{\infty}\!\!\!\!\!\!dP\ f^{W}(X,P,t)
	=\int_{-mc}^{mc}\!\!\!\!\!\!\!dP_0\sum_{\nu=-\infty}^{\infty}f^{W}(X,P_\nu(P_0),t).
	\label{relation between discrete and continuous momentum}
\end{align}

By differentiating Eq.~\eqref{Wigner distribution at local eq} with respect to $t$ and comparing the result with its spatial derivatives, we obtain the advection-diffusion equation~\eqref{advection-diffusion}.

Equation~\eqref{advection-diffusion} is formally equivalent to the Fokker-Planck equation for overdamped systems \cite{risken1996fokker}, which commonly describes directional Brownian motion driven by an external force or potential gradient \cite{risken1996fokker}, or classical Brownian ratchet systems, where the advection term arises from an asymmetric periodic potential combined with a time-periodic external field \cite{magnasco1993forced,astumian1994,astumian1997thermodynamics,reimann2002brownian}.
In such systems, the external influence is required to generate the advection term and to sustain directed transport, which is inevitably accompanied by energy dissipation, making the resulting advection an irreversible process.

In contrast, our 1D quantum system naturally exhibits an advection term in Eq.~\eqref{advection-diffusion} without external forces or potential gradients. This implies that directed transport in this system is sustained at local equilibrium without energy dissipation, and remarkably, the advection is a time-reversible effect.

The reversibility of the advection can be demonstrated as follows. Under time-reversal ${t \mapsto -t}$, the momentum is reversed ${P \mapsto -P}$, which in turn flips the sign of the sound velocity:
\begin{align}
    \sigma(P)&\xrightarrow{P \mapsto -P}
    \sigma(-P)=-\sigma(P).
    \label{advection symmetry 1}
\end{align}
This can be proven using Eqs.~\eqref{a paired subspace},\eqref{sound velocity 2}, and~\eqref{momentum dependency of the transport coefficients}.

This time-reversibility indicates that the initial conditions play a crucial role in the emergence of unidirectional transport. 
Specifically, if the initial state has an asymmetric momentum distribution, the paired subspaces ${\mathcal S}_{P_0}$ and ${\mathcal S}_{-P_0}$ carry different weights, leading to an asymmetric equilibrium momentum distribution across the entire momentum space. As a result, the total advection, integrated over all momentum space, becomes nonzero.
Conversely, if the initial state has a symmetric momentum distribution, the paired subspaces have equal weights, resulting in a symmetric momentum equilibrium distribution and, thus, zero total advection.

\begin{figure*}[htbp]
\centering{}
\includegraphics[scale=0.5]{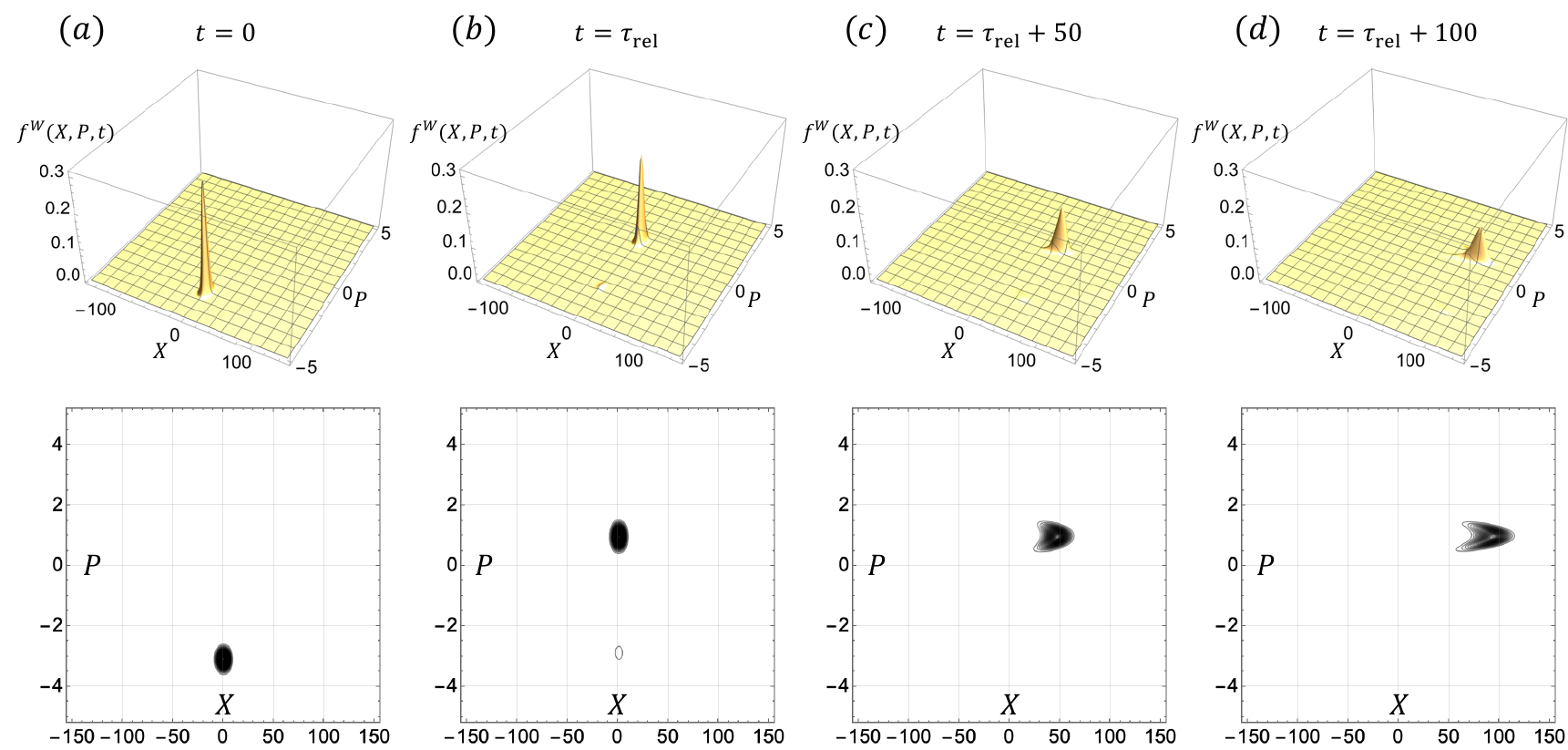}
\caption{
\label{ratchet_like_directional_motion}
Unidirectional transport of a one-dimensional quantum Brownian particle under the parameters chosen as $D^\ast/D_{\rm u}=1$.
All values are depicted in the units defined in Sec.~\ref{Model and Kinetic equation}. The temperature is set to $T=1$.
Unlike free particle propagation, even when a particle is initially given negative momentum, the particle propagation at local equilibrium $t\gtrsim\tau_{\mathrm{rel}}$ occurs in the positive spatial direction.
Note that during the relaxation period ($t=0$ to $t=\tau_{\mathrm{rel}}$), the momentum distribution of the particle relaxes to the thermal equilibrium in each subspace with the temperature of the phonons.
(a) Initial Wigner distribution at $t=0$ with a momentum peak at $P^{\prime}=-3.1$ with width $\Delta X = 3$. This peak momentum belongs to the subspace $P_0=0.9$, which means $P^{\prime}=P_2(0.9)=-3.1$.
(b) Wigner distribution immediately after reaching local equilibrium at $t=\tau_{\mathrm{rel}}$. Since momentum transfer occurs only within the subspace to which the initial momentum belongs, the equilibrium momentum distribution has a peak at $P_0=0.9$ with a small side peak at $P_2(0.9)=-3.1$. As the temperature increases, additional side peaks emerge at discrete momenta $P_{\nu}(0.9)$.
(c) and (d) are the Wigner distribution at $t=\tau_{\mathrm{rel}}+50$ and $t=\tau_{\mathrm{rel}}+100$, respectively. The distribution moves in the positive $X$ direction with the sound velocity $\sigma(0.9)>0$.
}
\end{figure*}

This behavior is reminiscent of free-particle propagation where the initial momentum asymmetry determines the direction of motion. However, unlike free-particle propagation, the emergence of the nonzero sound velocity, which is the effect proportional to $k$, in this system crucially depends on dissipation in the sense that it appears in the hydrodynamic regime after the momentum distribution has relaxed to the thermal equilibrium in each subspace. Moreover, since the directed motion appears at the local equilibrium, it is robust against noise. Nevertheless, the small diffusion effect proportional to $k^2$ causes the advection to gradually relax over time.

Figure~\ref{ratchet_like_directional_motion} illustrates the time evolution of the unidirectional transport of a 1D quantum Brownian particle, with an initial state prepared as a minimum uncertainty wave packet. 
The time evolution of the wave packet in this setup was previously derived (see Equation (88) in Ref.~\cite{nakade2020anomalous}).

As shown in Fig.~\ref{ratchet_like_directional_motion}, although the particle is initially given a negative momentum at $t=0$, the advection emerging at local equilibrium $t\gtrsim\tau_{\mathrm{rel}}$ proceeds in the positive spatial direction. 
This indicates that, in the 1D quantum system, the directed motion is not an underdamped motion where a particle's initial inertia persists, but rather a new structure emerging at local equilibrium.
Meanwhile, the wave packet gradually spreads out and its amplitude decreases due to the diffusion effect.

Since the advection term allows the particle to be transported in one direction with small spreading of the distribution due to diffusion, it enables more efficient information transmission than in the case without advection in Brownian computing. 
Importantly, our system exhibits the advection term without requiring external driving mechanisms or incurring energy costs.

\section{Advection-induced spatial contraction in 1D quantum Brownian motion
\label{Advection-induced spatial contraction in 1D quantum Brownian motion}
}

In the previous section, we have shown a mechanism by which a Brownian particle in the 1D quantum system follows the advection-diffusion equation without external force, and unidirectional transport is induced by the advection term. This results in a ratchet-like behavior without energy dissipation.

However, if a ratchet without energy dissipation exists, it could locally counteract diffusion. In particular, by installing two of ratchets in opposite directions, the system could lead to the contraction of the spatial distribution of the particle without paying any thermodynamic cost, which would appear to contradict the second law of thermodynamics.

In this section, we illustrate how, by setting specific initial conditions, the unidirectional transport in this 1D quantum system spontaneously leads to the contraction of the spatial distribution.
This phenomenon can be characterized by the phenomenologically defined diffusion coefficient, which is given by the time derivative of the mean-square displacement:
\begin{equation}
D^{(x)}(t)
:=
\frac{1}{2}\frac{d}{dt}
\langle(X-\langle X \rangle_{t})^{2}\rangle_{t},
\label{phenomenological diffusion coefficient}
\end{equation}
where the average $\langle \cdot \rangle_{t}$ is taken over the Wigner distribution function $f^{W}(X,P,t)$.

In this situation, the phenomenological diffusion coefficient $D^{(x)}(t)$ takes a negative value, meaning that the spatial distribution narrows over time, manifesting ``negative diffusion''.
At first glance, this spatial contraction might appear to contradict the second law of thermodynamics, as it suggests a spontaneous decrease in entropy associated with the uncertainty in the position of the particle. 
(Specifically, the uncertainty corresponds to the width of the probability distribution. For example, consider a Gaussian distribution $N(x; \mu_X,\sigma_X^2)$ with mean $\mu_X$ and standard deviation $\sigma_X$. Its entropy is given by $S^X=-\int dx \ N(x; \mu_X,\sigma_X^2)\ln N(x; \mu_X,\sigma_X^2)=\ln(2\pi e\sigma_X^2)/2$, which explicitly decreases as $\sigma_X$ decreases.)

However, as proven in our previous paper \cite{nakade2021anomalous}, the advection-diffusion equation~\eqref{advection-diffusion} always satisfies the H-theorem, ensuring that the second law of thermodynamics remains valid in this process.
For the reader's convenience, the proof of the H-theorem for Eq.~\eqref{advection-diffusion} is provided in Appendix~\ref{appendix H-theorem}.

In the next section, we will analyze the entropy balance of this situation to show the entropy decrease concerning the uncertainty in the position of the particle does not contradict the second law of thermodynamics.

In the latter part of this section, we analyze the spatial distribution contraction in terms of $D^{(x)}(t)$ to provide a detailed explanation for the mechanism.
The analytical solution of $D^{(x)}(t)$ for an arbitrary initial state is given [see Appendix~\ref{appendix D^x(t)}] by
\begin{align}
   D^{(x)}&(t\geq\tau_{{\mathrm{rel}}})
   =
      \nonumber \\
    &\bar{D} 
     +\Big\langle 
              \big(
              X-\langle X\rangle_{t=0}
              \big)
              \big(
              \sigma(P)-\langle \sigma(P) \rangle_{t=0}
              \big)
       \Big\rangle_{t=0}
       \nonumber\\
     &+t\ \Big\langle 
              \big(
                     \sigma(P)-\bar{\sigma}
              \big)^{2}
           \Big\rangle_{\rm eq},
     \label{analytic solution of D^x(t)}
\end{align}
where $\langle \cdot \rangle_{\rm eq}:=\langle \cdot \rangle_{t\geq\tau_{{\mathrm{rel}}}}$ is the average over the local equilibrium Wigner distribution $f^W(X,P,t\geq\tau_{\mathrm{rel}})$, and
\begin{align}
   \bar{D}:=\bigl\langle D(P)\bigr\rangle_{\rm eq},\ \   \bar{\sigma}:=\bigl\langle\sigma(P)\bigr\rangle_{\rm eq},
   \label{average of transport coefficients}
\end{align}
are the averaged transport coefficients.

A notable feature of this system is that, in addition to the first term of Eq.\eqref{analytic solution of D^x(t)} representing diffusion, a second term that depends on the initial conditions and a third term that is linear in time also appear. These additional terms arise from the momentum dependence of the sound velocity $\sigma(P)$, and, therefore, represent the reversible phase mixing effects \cite{nakade2020anomalous}.

Among these, the second term is the only term that can take negative values, as the first term is an average of the diffusion coefficient $D(P)$, which is always positive, and the third term is an average of a squared quantity. 

The negativity of the second term depends on the initial conditions. This dependence arises because momentum relaxation occurs independently within each subspace ${\mathcal S}_{P_0}$, meaning that the equilibrium average retains a memory of the initial distribution. The derivation in Appendix~\ref{appendix D^x(t)} clarifies this point in detail.

\subsection{Initial conditions for spatial contraction: Nonfactorizable Gaussian state
\label{Initial conditions for spatial contraction: Nonfactorizable Gaussian state}}

To investigate the contraction of the spatial distribution in this system, we employ a Gaussian wave packet as the initial pure state. Gaussian pure states belong to a class of quantum states whose Wigner distribution functions are strictly nonnegative \cite{hudson1974wigner}. Moreover, in our system, the Wigner distribution function remains nonnegative at local equilibrium for such an initial state, and no negative regions appear thereafter.

Thus, starting from a Gaussian pure state eliminates the possibility that the spatial contraction could stem from “negative probability” regions in the Wigner distribution. Hence, we can rule out the quantum interference artifacts as the origin of the spatial contraction.

Under this condition, the Wigner distribution function can be regarded as a joint probability distribution over coordinate and momentum variables \cite{hudson1974wigner}. Consequently, the functional $-\int \!dXdP\  f^W\ln f^W$ coincides with the standard non-equilibrium entropy, allowing us to analyze the entropy balance in a classical probabilistic sense. This analysis will be presented in Sec.~\ref{Entropy balance during the spatial distribution contraction}.


Note that although we choose a pure initial state for the particle, the phonon bath is assumed to be in thermal equilibrium, and hence the total system is in a mixed state from the outset.
Importantly, the mechanism of spatial contraction does not require the initial state for the particle to be pure.
In practical implementations, mixed states could also lead to spatial contraction.

To induce spatial contraction, the initial state must be arranged so that different spatial regions exhibit opposite signs of the sound velocity $\sigma(P)$. Satisfying this requirement necessitates that different spatial regions must carry different momenta, meaning that there is a non-zero correlation between coordinates and momenta. Once such a correlation emerges, the distribution can no longer be factorized into independent functions of $X$ and $P$.

For this purpose, we use a nonfactorizable Gaussian pure state whose Wigner distribution function is given as follows:
\begin{align}
  &f^{W}(X,P,t=0)
  \nonumber\\
  &=
  \frac{1}{\pi\hbar}
    \exp
      \left[
            -\frac{(P-P^{\prime})^{2}}
                  {2(\Delta P)^{2}}
            -\frac{(X-X^{\prime}-\alpha P)^{2}}
                  {2(\widetilde{\Delta X})^2}
      \right],
      \label{initial squeezed Gaussian}
\end{align}
where the parameter $\Delta P$ represents the standard deviation of $P$, and $\widetilde{\Delta X}$ is defined by the relation
\begin{equation}
  \widetilde{\Delta X}\cdot\Delta P=\frac{\hbar}{2}.
  \label{deceptive minimum uncertainty}
\end{equation}
The standard deviation $\Delta X$ of $X$ is given by
\begin{equation}
   \Delta X=\sqrt{(\widetilde{\Delta X})^{2}+\alpha^2 (\Delta P)^2},
  \label{widetildeDeltaX}
\end{equation}
and therefore the state is not a minimum uncertainty state unless $\alpha=0$.
The parameters $X^{\prime}$ and $P^{\prime}$ represent the peak positions of the wave packet.
The parameter $\alpha$ has the dimension $X/P$ and introduces a momentum-dependent shift in the spatial distribution.

We can verify that the state~\eqref{initial squeezed Gaussian} is a pure state under the condition~\eqref{deceptive minimum uncertainty}.
By applying a Fourier transformation to Eq.~\eqref{initial squeezed Gaussian}, we obtain the corresponding reduced density matrix:
\begin{align}
 &f_{k}(P,t=0) 
\nonumber \\
 &=
  \frac{1}
       {\sqrt{2\pi(\Delta P)^{2}}}
    \exp
      \Biggl[
            -\frac{(P-P^{\prime})^{2}}{2(\Delta P)^{2}}
    \nonumber \\
    &\ \ \ \ \ \ \ \ \ \ \ \ \ \ \ \ \ \ \ \ \ \ \ \ \ \ 
            -\frac{(\widetilde{\Delta X})^{2}}{2}
               k^{2}
            -i(X^{\prime}+\alpha P)k
      \Biggr],
      \label{fk for initial squeezed Gaussian}
\end{align}
which can be factorized with the condition~\eqref{deceptive minimum uncertainty} into the following form:
\begin{equation}
  f_{k}(P,t=0) 
  =
  \psi_{0}(P+\frac{\hbar k}{2})
    \psi_{0}^{\ast}(P-\frac{\hbar k}{2}),
      \label{eq:fk factorized by wave fn}
\end{equation}
where the wave function $\psi_{0}$ is written as
\begin{align}
  &\psi_{0}(P)
  :=
  \left(P\middle|\psi_{0}\right)
  \nonumber \\
  & =
    \left[\frac{1}{2\pi(\Delta P)^2}\right]^{\frac{1}{4}}
    \exp\left[
              -\frac{(P-P^{\prime})^2}{4(\Delta P)^2}
              -i\frac{P}{\hbar}X^{\prime}
              -\frac{i}{\hbar}\frac{\alpha}{2}P^2
        \right].
   \label{psi_{0}(P)}
\end{align}
Here, the round bracket in the expression is normalized by the delta function as defined in Eq.~\eqref{delta normalize}.

In the following, we describe the correlation between $X$ and $P$ in the initial distribution~\eqref{initial squeezed Gaussian}, which plays the essential role for spatial contraction.
The presence of the $-\alpha P$ term in the exponent in Eq.~\eqref{initial squeezed Gaussian} prevents the distribution from being factored into separate functions of $X$ and $P$, which establishes the non-zero correlation between $X$ and $P$.
This correlation can be quantified by the correlation coefficient:
\begin{equation}
   r:=\frac{\mathrm{Cov}(X,P)}
              {
               \sqrt{
                       \mathrm{Var}(X)
                       \mathrm{Var}(P)
                       }
               }
     =\frac{\alpha \Delta P}{\Delta X},
     \label{correlation coefficient}
\end{equation}
where ${\rm Var}(X)$ and ${\rm Var}(P)$ are the variance of $X$ and $P$, respectively, and ${\mathrm{Cov}(X,P)}$ is the covariance between $X$ and $P$:
\begin{align}
\mathrm{Var}(X) &:= (\Delta X)^2 := \langle X^2 \rangle - \langle X \rangle^2,
\\
\mathrm{Var}(P) &:= (\Delta P)^2 := \langle P^2 \rangle - \langle P \rangle^2,
\\
\mathrm{Cov}(X,P) &:= \tfrac{1}{2} \langle XP + PX \rangle - \langle X \rangle \langle P \rangle=\alpha (\Delta P)^2.
\end{align}
Hence, the sign of the correlation coefficient is determined by the parameter $\alpha$.

Next, we demonstrate that spatial contraction arises when this nonfactorizable Gaussian pure state is given as an initial state. 
However, the following argument does not necessarily have to satisfy Eq.~\eqref{deceptive minimum uncertainty}, so it is valid even in mixed states.
The Wigner distribution function under local equilibrium for the given initial state can be obtained by substituting 
Eq.~\eqref{fk for initial squeezed Gaussian} into Eqs.~\eqref{Wigner distribution at local eq} and~\eqref{chi at local eq}, yielding:
\begin{widetext}
\begin{align}
  f^{W}(X,P_{\nu},t\gtrsim\tau_{\mathrm{rel}})
  = & 
  \ \varphi_{P_{0}}^{ {\mathrm{eq}}}(P_{\nu})
  \sum_{\mu=-\infty}^{\infty}
    \frac{1}{\sqrt{2\pi(\Delta P)^{2}}}
      \exp
      \left[
            -\frac{(P_{\mu}-P^{\prime})^{2}}{2(\Delta P)^{2}}
      \right]
      \nonumber
      \\
   &\times
      \frac{1}
           {\sqrt{2\pi\{(\widetilde{\Delta X})^{2}+2D(P_{0})t\}}}
      \exp
      \left[
            -\frac{(X-X^{\prime}-\alpha P_{\mu}-\sigma(P_{0})t)^{2}}
                  {2\{(\widetilde{\Delta X})^{2}+2D(P_{0})t\}}
      \right].
      \label{chi of squeezed Gaussian at local eq}
\end{align}
\end{widetext}
Here, the summation over $\mu$ runs over all discrete momenta $P_\mu$ within the subspace ${\mathcal S}_{P_0}$ to which the discrete momentum $P_{\nu}$ belongs.
Note that this function remains strictly nonnegative and that no negative regions appear at any $t\gtrsim \tau_\mathrm{rel}$.

By substituting $t=0$ into Eq.~\eqref{chi of squeezed Gaussian at local eq}, we can see the form of the initial distribution given in Eq.~\eqref{initial squeezed Gaussian}, indicating that the correlation between coordinate and momentum present in the initial state remains after the system reaches local equilibrium. 
In systems with two or more dimensions, relaxation occurs across the entire momentum space, and this summation over $\mu$ is replaced by an integral over the entire momentum space. 
As a result, the correlation between coordinate and momentum in the initial state vanishes at local equilibrium, and the distribution takes a form that is completely separable in the variables of coordinate $X$ and momentum $P$.
Hence, the persistence of correlation is a distinctive feature of the 1D system.

\begin{figure}[htbp]
\centering{}
\includegraphics[scale=0.5]{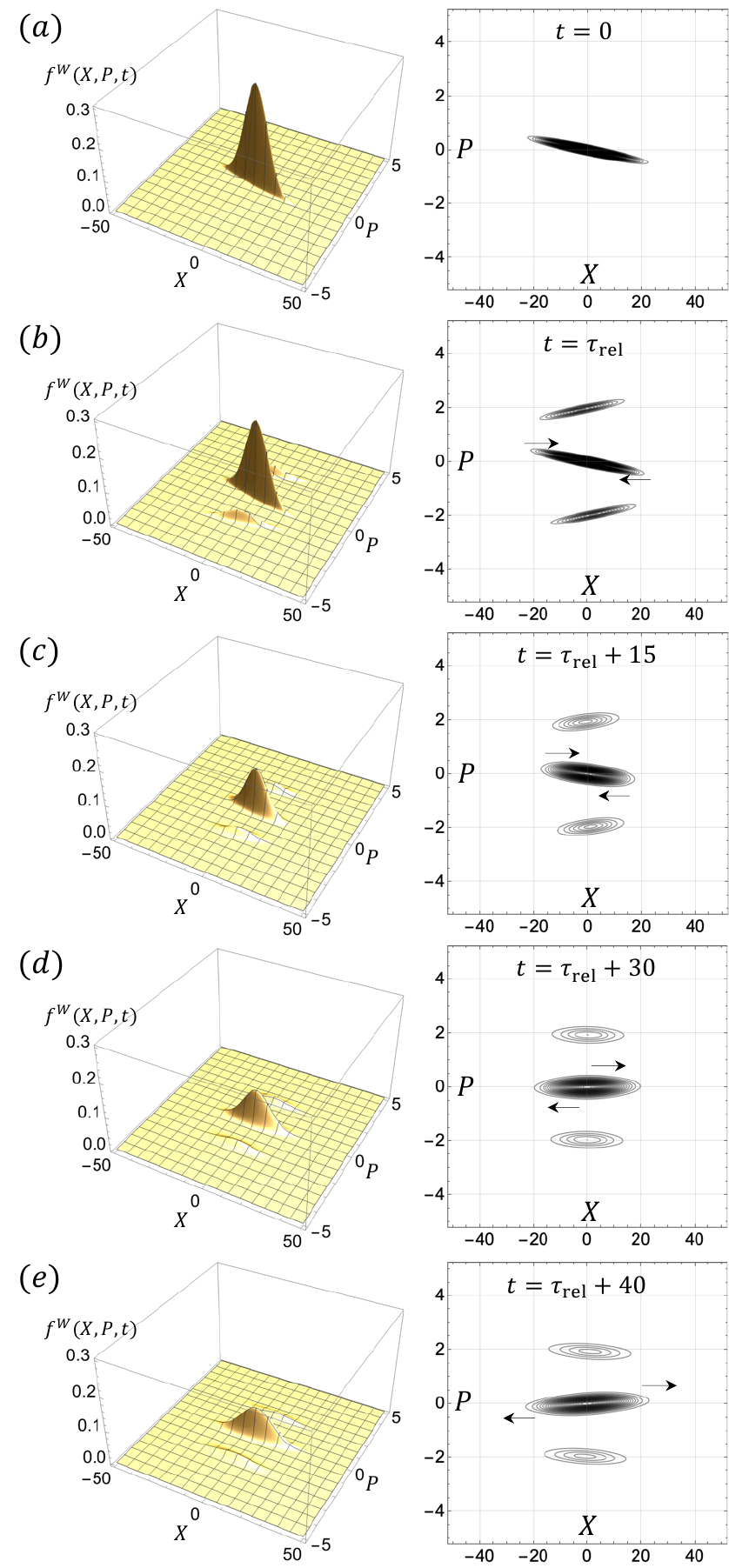}
\caption{
\label{apparently_negative_diffusion}
Apparent negative diffusion with a single nonfactorizable Gaussian wave packet under the parameters chosen as $D^\ast/D_{\mathrm u}=1$. All values are depicted in the units as described in Sec.~\ref{Model and Kinetic equation}. 
The left side shows bird's-eye views of the Wigner distribution function, and the right side presents contour plots. 
The arrows in the contour plots indicate the direction of momentum-dependent advection. 
The temperature is set to $T=1$ in the aforementioned units.
(a) Initial distribution at $t=0$ with the negative shift parameter $\alpha=-40$ (in units of $x_{\mathrm u}/p_{\mathrm u}$).  The peak of the Gaussian is set at the origin $(X^{\prime}, P^{\prime})=(0,0)$ with its width $\widetilde{\Delta X} = 3$.
(b) Wigner distribution immediately after reaching local equilibrium at $t=\tau_\mathrm{rel}$. Since the initial momentum peak is set at $P_0=P^{\prime}=0$, side peaks appear at $P_1(P_0)=2$ and $P_{-1}(P_0)=-2$. 
(c) Wigner distribution at $t=\tau_\mathrm{rel}+15$. Although diffusion spreads the wave packet, advection dominates, leading to the contraction of the wave packet in the spatial direction.
(d) Wigner distribution at $t=\tau_\mathrm{rel}+30$. The tilt in the contour of the distribution has almost vanished due to advection. 
(e) Wigner distribution at $t=\tau_\mathrm{rel}+40$. At this stage, both the diffusion and the advection contribute to the spreading of the wave packet.
}
\end{figure}

The time evolution of the Wigner distribution function~\eqref{chi of squeezed Gaussian at local eq} is depicted in Fig.~\ref{apparently_negative_diffusion}. 
Here, we set the initial shift parameter $\alpha$ to be negative, and set the peak of the Gaussian at the origin $(X^{\prime}, P^{\prime})=(0,0)$.
At the initial state $t=0$, the distribution exhibits a negative correlation between coordinate $X$ and momentum $P$, as shown by the tilt of the elliptical contour in the $X$-$P$ plane. 

After reaching local equilibrium at $t=\tau_\mathrm{rel}$, the distribution splits into a central peak at $P_0=P^{\prime}$ and side peaks at $P_{\pm 1}$ along the momentum axis. This splitting occurs because momentum transitions are restricted to the subspaces associated with the initial momentum distribution.

As shown in Fig.~\ref{apparently_negative_diffusion}~(b), while the main peak at $P_0$ exhibits a negative correlation, the side peaks at $P_{\pm 1}$ (which also belong to the same subspace) show a positive correlation.
This inversion of correlation occurs because the sign of the $P_0$ dependence of $P_{\mu}$ flips with $\mu$:
\begin{equation}
   \frac{d P_{\mu}}{d P_0}=(-1)^{\mu}.
\end{equation}
Importantly, despite the flip in $X$-$P_{\mu}$ correlation, all peaks at $P_{\mu}$ share the same $X$-$P_0$ correlation (negative in the present case) originating from the initial state.
Note that a necessary condition for spatial contraction is that the distribution exhibits a negative $X$-$P_0$ correlation.

As is evident from the contour plots, the wave packet is contracting in the spatial direction from $t=\tau_\mathrm{rel}$ to $t=\tau_\mathrm{rel}+15$. This occurs because the sound velocity $\sigma(P)$ has opposite signs at opposite sign momenta $P$ and $-P$, as shown in Eq.~\eqref{advection symmetry 1}. Consequently, the distortion in the distribution, which reflects the negative correlation, gradually disappears.

At a later time, $t=\tau_\mathrm{rel}+30$, the distortion reflecting the negative correlation almost disappears, and by $t=\tau_\mathrm{rel}+40$, the distribution is instead distorted to exhibit positive correlation. In other words, the advection acts to contract the wave packet only until the negative correlation created in the initial state disappears. Beyond this time domain, the advection begins to work in the opposite direction, spreading the wave packet. It should also be noted that at all times, the diffusion effect works to spread the wave packet, gradually lowering its amplitude.

\subsection{Analysis of the phenomenological diffusion coefficient
\label{Analysis of the phenomenological diffusion coefficient}}

Next, we analyze the phenomenological diffusion coefficient $D^{(x)}(t)$ in the situation where apparent negative diffusion is occurring. 

First, let us consider the diffusion term represented by the first term in the analytical solution~\eqref{analytic solution of D^x(t)}:
\begin{align}
	\bar{D}
	&:=
	\left\langle D(P)\right\rangle _{t=\tau_{\mathrm{rel}}}
	\nonumber\\
	&=
	\int_{-\infty}^{\infty}\!\!\!\!\!\!\!dX
	\int_{-\infty}^{\infty}\!\!\!\!\!\!\!dP\
	 D(P)f^{W}(X,P,t=\tau_{\mathrm{rel}})>0.
	\label{barD definition}
\end{align}
The reason this inequality holds is that the microscopic diffusion coefficient is always positive, $D(P) > 0$, and the Wigner distribution function integrated over $X$ represents the true probability distribution for momentum:
\begin{align}
	\varphi(P,t):= 
	\int_{-\infty}^{\infty}\!\!\!\!\!\!\!dX\ f^{W}(X,P,t)\geq0.
	\label{eq:varphi_P(t)}
\end{align}
Therefore, for any initial condition, the diffusion always contributes to the spreading of the wave packet.

Next, we focus on the second term, which depends on the initial distribution: 
\begin{align}
	\Big\langle 
              \big(
                X-\langle X\rangle_{t=0}
              \big)
              \big(
                \sigma(P)-\langle \sigma(P) \rangle_{t=0}
              \big)
       \Big\rangle_{t=0} 
	\nonumber \\
	=
	\big\langle 
	            X \sigma(P) 
	\big\rangle_{t=0}
	-
	\big\langle 
	            X 
	\big\rangle_{t=0}
	\big\langle
	           \sigma(P)
	\big\rangle_{t=0}.
	\label{new term for Dx(t)}
\end{align}
This term arises due to the momentum dependence of the sound velocity $\sigma(P)$. Therefore, it represents a reversible phase mixing effect.
This term vanishes if the initial Wigner distribution function is factorizable into independent functions of $X$ and $P$ as:
\begin{equation}
	f^{W}(X,P,t=0)=\chi(X) \varphi(P).
	\label{separable fW(0)}
\end{equation}

The second term~\eqref{new term for Dx(t)} can also take either positive or negative values since the sound velocity $\sigma(P)$ can take both positive and negative values.
When this term becomes negative, the reversible phase mixing effect contributes to the contraction of spatial distribution, competing with the irreversible diffusion effect that spreads the distribution. 
Especially, if the condition
\begin{align}
	\bar{D}
	+
	\Big\langle 
              \big(
                X-\langle X\rangle_{t=0}
              \big)
              \big(
                \sigma(P)-\langle \sigma(P) \rangle_{t=0}
              \big)
       \Big\rangle_{t=0} 
       <0,
\end{align}
holds, then apparent negative diffusion occurs. However, since this negative contribution of the phase mixing is a reversible effect, it does not contribute to the decrease of entropy.

Finally, let us focus on the third term in Eq.~\eqref{analytic solution of D^x(t)}, which depends linearly on time $t$. This term also arises because the sound velocity $\sigma(P)$ depends on momentum, reflecting the effect of phase mixing. However, unlike the second term, this term always contributes to the spreading of the wave packet, because the condition
\begin{equation}
   t\ \Big\langle 
              \big(
                     \sigma(P)-\bar{\sigma}
              \big)^{2}
           \Big\rangle_ {\mathrm{eq}}>0,
   \label{t linear term of D^x}
\end{equation}
holds. Furthermore, the third term increases linearly with time, leading to anomalous diffusion. The anomalous diffusion caused by this phase mixing effect is discussed in detail in \cite{nakade2020anomalous}.

\begin{figure}[t]
\centering{}
\includegraphics[scale=0.6]{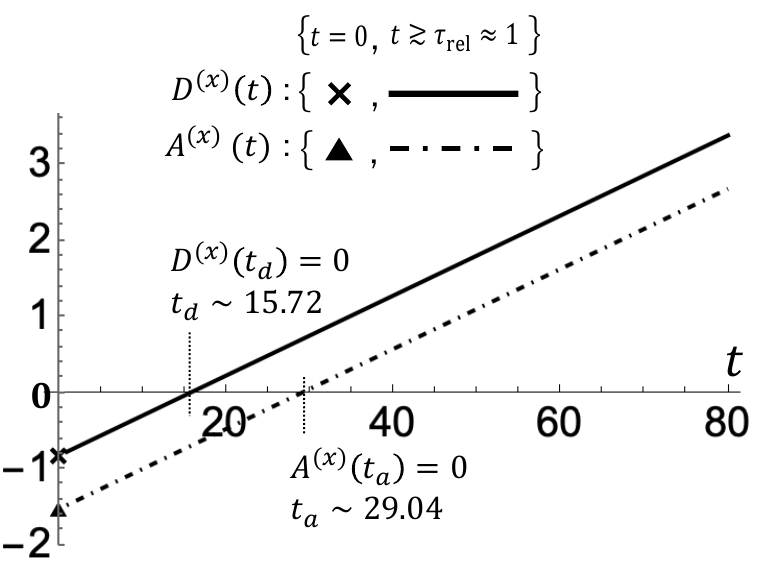}
\caption{
\label{fig_phenomenological_diffusion_constant}
Time evolution of the phenomenological diffusion coefficient $D^{(x)}(t)$ and the effect excluding the diffusion effect $A^{(x)}(t) = D^{(x)}(t) - \bar{D}$ in the situation depicted in Fig.~\ref{apparently_negative_diffusion}. 
The relaxation time (in units of $t_{\mathrm u}$) is $\tau_\mathrm{rel} \approx 1$ under the parameters chosen as $D^\ast=D_{\mathrm u}$. 
The cross marks represent $D^{(x)}(0)$, the solid line shows $D^{(x)}(t \gtrsim \tau_\mathrm{rel})$, the triangle marks indicate $A^{(x)}(0)$, and the dashed-dotted line depicts $A^{(x)}(t \gtrsim \tau_\mathrm{rel})$. 
The time $t_d$ satisfies $D^{(x)}(t_d) = 0$, which corresponds to the threshold time, at which contraction turns into expansion, resulting in the minimum width of the spatial distribution. 
Additionally, the time $t_a$ satisfies $A^{(x)}(t_a) = 0$, at which the negative tilt in the contour of the distribution turns into a positive one.
}
\end{figure}

Figure~\ref{fig_phenomenological_diffusion_constant} shows the time evolution of the phenomenological diffusion coefficient $D^{(x)}(t)$ as the solid line, corresponding to the situation depicted in Fig.~\ref{apparently_negative_diffusion}. 
Initially, $D^{(x)}(t)$ takes negative values. This indicates that the reversible phase mixing effect, which provides the negative contribution, overwhelms the positive contribution from diffusion, leading to the contraction of the wave packet in the spatial direction at that time.
However, due to the positive phase mixing effect~\eqref{t linear term of D^x}, which provides the positive time-linear contribution,  $D^{(x)}(t)$ turns to be positive and the wave packet begins to expand. 

The threshold time $t_d$, at which $D^{(x)}(t_d)=0$ satisfies and contraction turns into expansion, is given by
\begin{equation}
   t_d =
   \frac
   {-
	\Big\langle 
              \big(
                X-\langle X\rangle_{t=0}
              \big)
              \big(
                \sigma(P)-\langle \sigma(P) \rangle_{t=0}
              \big)
       \Big\rangle_{t=0}
   -\bar{D}
   }
   {
   \ \Big\langle 
              \big(
                     \sigma(P)-\bar{\sigma}
              \big)^{2}
           \Big\rangle_ {\mathrm{eq}}
     }.
     \label{td}
\end{equation}
In Fig.~\ref{fig_phenomenological_diffusion_constant}, $t_d \sim 15.72$ with the time unit $t_{\mathrm u}=\hbar/mc^2$, which approximately corresponds to the situation at $t=\tau_\mathrm{rel}+15$ in Fig.~\ref{apparently_negative_diffusion}.  
One can see that at $t=\tau_\mathrm{rel}+15$ in Fig.~\ref{fig_phenomenological_diffusion_constant}, the tilt in the contour of the distribution has not yet disappeared. Nevertheless, the phenomenological diffusion coefficient $D^{(x)}(t)$ turns to positive, indicating the beginning of wave packet expansion.

To understand this behavior, we decompose $D^{(x)}(t)$ into its reversible and irreversible contributions by subtracting the irreversible diffusion term $\bar{D}$, leading to the definition:
\begin{align}
   A^{(x)}(t): =D^{(x)}(t)-\bar{D},
\end{align}
This represents the diffusion-free component that arises solely from reversible phase mixing effects.

The time evolution of $A^{(x)}(t)$ is plotted as the dashed-dotted line in Fig.~\ref{fig_phenomenological_diffusion_constant}.
The threshold time $t_a$, at which $ A^{(x)}(t_a)=0$, is greater than $t_d$ and is given by
\begin{align}
   t_a =
   -\frac
   {
	\Big\langle 
              \big(
                X-\langle X\rangle_{t=0}
              \big)
              \big(
                \sigma(P)-\langle \sigma(P) \rangle_{t=0}
              \big)
       \Big\rangle_{t=0}
   }
   {
   \ \Big\langle 
              \big(
                     \sigma(P)-\bar{\sigma}
              \big)^{2}
           \Big\rangle_ {\mathrm{eq}}
     }.
\end{align}
In Fig.~\ref{fig_phenomenological_diffusion_constant}, $t_a\sim 29.04$, which approximately corresponds to the situation at $t=\tau_\mathrm{rel}+30$ in Fig.~\ref{apparently_negative_diffusion}. At this point, the tilt in the contour of the distribution has nearly disappeared due to advection.
For $t > t_a$, the tilt gradually reappears in the opposite direction as advection continues.

In summary, the initial conditions create a competition between reversible advection, which contracts the distribution, and irreversible diffusion, which always expands the distribution. When advection is sufficiently strong compared to diffusion, the spatial distribution contracts. 
The contribution of advection changes over time. Initially, advection counteracts diffusion, leading to the spatial contraction. However, beyond a finite time $t_a$ (which depends on the initial shift parameter $\alpha$), the advection effect starts to contribute to the expansion of the distribution. The minimum width of the spatial distribution is reached at a time $t_d<t_a$, due to the interplay between advection and diffusion.

As the diffusion effect represented by $\bar{D}$ strengthens with increasing temperature, it gradually reduces the time $t_d$ and the time interval
\begin{equation}
   \tau_\mathrm{rel}\lesssim t\leq t_d,
   \label{time interval}
\end{equation}
where the spatial contraction can be observed, diminishes. Above a threshold temperature, the difusion dominates and the system does not exhibit the spatial contraction.
In Fig.~\ref{fig_parameter_dependence_of_spatial_contraction} in Appendix~\ref{appendix threshold temperature}, we show this threshold temperature, using parameter values relevant to the protein molecular chain.

\section{
Entropy balance during the spatial distribution contraction
\label{Entropy balance during the spatial distribution contraction}
}

The previous section illustrates a situation where the spatial distribution of a Brownian particle in one-dimensional quantum system contracts without releasing heat to the environment, and the H-theorem still holds. 
However, what compensates for the decrease in entropy associated with this contraction of the spatial distribution?
In this section, we answer this question by analyzing the entropy balance in the situation.

First, we discuss the method of defining entropy using the Wigner distribution function. 
In general, the Wigner distribution function is a quasi-joint probability distribution over coordinate and momentum, which, due to the uncertainty principle, can take negative values.
Therefore, if one defines non-equilibrium entropy in the form of $-\int \!dXdP\  f^W\ln f^W$, the entropy value may have an imaginary part, and it is problematic. 

However, in the situation of Sec.~\ref{Advection-induced spatial contraction in 1D quantum Brownian motion}, where the initial state is given as a single Gaussian wave packet, the Wigner distribution function remains non-negative throughout its time evolution, and it becomes a joint probability distribution over coordinate and momentum variables \cite{hudson1974wigner}. Therefore, in this specific case, we can define entropy using the Wigner distribution function as follows.
\begin{equation}
  S(t)
  :=
  -\int dX\int dP\ 
    f^{W}(X,P,t)
    \ln\{
        f^{W}(X,P,t)/\frac{1}{\pi\hbar}
       \}.
  \label{eq:S(t)}
\end{equation}
Here, the factor $1/\pi\hbar$ inside the logarithm is to make the expression dimensionless. Since the Wigner distribution function has upper and lower bounds \cite{leonhardt1995measuring},
\begin{equation}
   |f^{W}(X,P,t)| \leq \frac{1}{\pi\hbar},
   \label{below and upper bound of Wigner function}
\end{equation}
the factor $1/\pi\hbar$ guarantees that the entropy $S(t)$ takes positive values.

The entropy $S(t)$ increases monotonically:
\begin{align}
	 \frac{d}{dt}
          S(t)\geq 0. 
	\label{d/dt S}
\end{align}
The proof is done by imposing the condition $C = 0$ (See Eq.~\eqref{bottom-up function}) and $f^W(X,P,t) \geq 0$ in Appendix~\ref{appendix H-theorem}.

This entropy can be identically decomposed as follows:
\begin{equation}
  S(t)
  =
  S^{X}(t)+S^{P}(t)-S_{I}^{X:P}(t), 
  \label{S^{X}(t)+S^{P}(t)-S_{I}^{X:P}(t)}
\end{equation}
where
\begin{align}
  S^{X}(t) & 
  :=
  -\int \!dX\ 
    \chi(X,t)
      \ln\{
           \chi(X,t)/\frac{1}
                          {\Delta_{\chi}}
         \},\label{S^{X}(t)}
         \\
  S^{P}(t) & 
  :=
  -\int \!dP\ 
    \varphi(P,t)
      \ln\{
           \varphi(P,t)/\frac{1}
                             {\Delta_{\varphi}}
         \},\label{S^{P}(t)}
         \\
  S_{I}^{X:P}(t) & 
  :=
  \int \!dX \!
  \int \!dP\ 
    f^{W}(X,P,t)
      \ln
          \left\{
          \frac{f^{W}(X,P,t)}{\chi(X,t)\varphi(P,t)}
         \right\},
  \label{S_{I}^{X:P}(t)}
\end{align}
and $\chi(X,t)$ and $\varphi(P,t)$ are the marginal probability distributions for coordinate and momentum, respectively:
\begin{align}
  \chi(X,t) & 
  =
  \int dP\ f^{W}(X,P,t),\\
  \varphi(P,t) & 
  =
  \int dX\ f^{W}(X,P,t).
\end{align}
The factors $\Delta_{\chi}$ and $\Delta_{\varphi}$ inside the logarithms in Eqs.~\eqref{S^{X}(t)} and~\eqref{S^{P}(t)} are used to make the expressions dimensionless, and they satisfy 
\begin{equation}
      \Delta_{\chi}\cdot\Delta_{\varphi}=\pi\hbar.
\end{equation}
Here, we choose
\begin{equation}
      \Delta_{\chi}=\sqrt{2\pi(\widetilde{\Delta X})^{2}},\ \ \Delta_{\varphi}=\sqrt{2\pi(\Delta P)^{2}}.
\end{equation}
Since these factors, $\Delta_{\chi}$ and $\Delta_{\varphi}$, do not contribute to the time evolution of each entropy term, they are not essential for the subsequent discussion.

The entropy of the spatial distribution, $S^{X}(t)$, represents the uncertainty in the coordinate of the particle, while the entropy of the momentum distribution, $S^{P}(t)$, represents the uncertainty in the particle's momentum. 
Additionally, $S_{I}^{X:P}(t)$ represents the mutual information between coordinate and momentum  \cite{shannon1948mathematical}. 
This quantity is defined as the relative entropy (Kullback-Leibler divergence) \cite{cover1999elements} between the joint probability distribution $f^{W}(X, P,t)$ and the product of the marginal distributions $\chi(X,t)\varphi(P,t)$. 
The mutual information $S_{I}^{X:P}(t)$ quantifies the degree of correlation between $X$ and $P$. 
It can be proven that $S_{I}^{X:P}(t)$ is always non-negative \cite{shannon1948mathematical}, and a decrease in $S_{I}^{X:P}(t)$ contributes to an increase in the overall entropy $S(t)$.

The identity (\ref{S^{X}(t)+S^{P}(t)-S_{I}^{X:P}(t)}) suggests the possibility that the spatial distribution entropy $S^{X}(t)$ can decrease, while the total entropy $S(t)$ still satisfies the second law of thermodynamics~\eqref{d/dt S}, with the decrease compensated by the reduction in the mutual information $S_{I}^{X:P}(t)$.

However, in classical Brownian motion, it is assumed that the momentum distribution relaxes to the Maxwellian distribution, independent of the spatial distribution. As a result, the mutual information decreases to $S_{I}^{X:P}(t)=0$, leaving no room for a decrease at local equilibrium, and thus the above entropy balance is not possible.

\begin{widetext}
On the other hand, in the one-dimensional quantum Brownian motion, the mutual information remains greater than zero even after reaching local equilibrium since the joint probability distribution at local equilibrium is non-separable between $X$ and $P$, as given in Eqs.~\eqref{Wigner distribution at local eq} and~\eqref{chi at local eq}:
\begin{equation}
   S_{I}^{X:P}(t\gtrsim\tau_\mathrm{rel})
  =
  \int dX\int^{mc}_{-mc}\!\!\!\!\! dP_0 \sum_{\nu=-\infty}^{\infty}\ 
     \chi_{P_{0}}(X,t)
     \varphi_{P_{0}}^{ {\mathrm{eq}}}(P_{\nu})
       \ln\{
          \frac{ \chi_{P_{0}}(X,t)
                 \varphi_{P_{0}}^{ {\mathrm{eq}}}(P_{\nu})
                }
                {\chi(X,t)\varphi(P_\nu,t)}
          \}>0.
    \label{relative entropy at local eq}
\end{equation}
Here, the relation between $\chi(X,t)$ and $\chi_{P_{0}}(X,t)$, and  between $\varphi(P_{\nu},t)$ and $\varphi_{P_{0}}^{ {\mathrm{eq}}}(P_{\nu})$ are as follows [with use of relation~\eqref{relation between discrete and continuous momentum}]: 
\begin{align}
  \chi(X,t\gtrsim\tau_\mathrm{rel}) & 
  =
  \int_{-mc}^{mc}\!\!\!\!\!\!dP_{0}
    \sum_{\nu=-\infty}^{\infty}
      \chi_{P_{0}}(X,t)
      \varphi_{P_{0}}^{ {\mathrm{eq}}}(P_{\nu})
  =
  \int_{-mc}^{mc}\!\!\!\!\!\!dP_{0}\ 
    \chi_{P_{0}}(X,t),
    \label{chi_X}
    \\
  \varphi(P_{\nu},t\gtrsim\tau_\mathrm{rel}) & 
  =
  \int dX\ 
    \chi_{P_{0}}(X,t)
    \varphi_{P_{0}}^{ {\mathrm{eq}}}(P_{\nu})
  =
  \varphi^{{\mathrm{init}}}(P_{0})
    \varphi_{P_{0}}^{ {\mathrm{eq}}}(P_{\nu}),
    \label{phi_P}
\end{align}
\end{widetext}
where
\begin{equation}
  \varphi^{{\mathrm{init}}}(P_{0})
  :=
  \int dX\ 
    \chi_{P_{0}}(X,t)
  =
  \sum_{\mu=-\infty}^{\infty}
    \varphi(P_{\mu}(P_0),0), 
    \label{varphiinit}
\end{equation}
is a conserved quantity in each momentum subspace ${\mathcal S}_{P_0}$. 
[See Eq.~\eqref{proof of varphiinit} in Appendix~\ref{appendix D^x(t)} for the derivation of~\eqref{varphiinit}.]

The time derivatives of the entropies~\eqref{S^{X}(t)},~\eqref{S^{P}(t)}, and~\eqref{S_{I}^{X:P}(t)} are written as follows:
\begin{align}
  \frac{d}{dt}S^{X}(t) 
  = & 
  -\frac{d}{dt}
    \int dX\ \chi(X,t)
    \ln\chi(X,t),
    \label{dSx}\\
  \frac{d}{dt}
    S^{P}(t) 
  = & \ 0,\\
  \frac{d}{dt}
    S_{I}^{X:P}(t) 
  = & 
  -\frac{d}{dt}
    \int dX\ 
      \chi(X,t)
        \ln\chi(X,t)
   \nonumber\\
  &+\frac{d}{dt}
    \int_{-mc}^{mc}\!\!\!\!\!\!dP_{0}
    \!
    \int_{-\infty}^{\infty}\!\!\!\!\!\!dX\ 
      \chi_{P_{0}}(X,t)
        \ln\chi_{P_{0}}(X,t).
    \label{dSI}
\end{align}
Thus, the time derivatives of the total entropy $S(t)$ is as follows:
\begin{align}
  \frac{d}{dt}S(t) & 
  =
  \frac{d}{dt}
    S^{X}(t)
  -\frac{d}{dt}
    S_{I}^{X:P}(t)
    \nonumber\\
 & =
   -\frac{d}{dt}
     \int_{-mc}^{mc}\!\!\!\!\!\!dP_{0}
     \int_{-\infty}^{\infty}\!\!\!\!\!\!dX\ 
       \chi_{P_{0}}(X,t)
         \ln\chi_{P_{0}}(X,t)
     \nonumber\\
 &=
    -\int_{-mc}^{mc}\!\!\!\!\!\!\!dP_0\ D(P_0)
    \!
     \int_{-\infty}^{\infty}\!\!\!\!\!\!\!dX
	  \left[
		 \frac{
		         \{\frac{\partial}{\partial X}\chi_{P_{0}}(X,t)\}^{2}
		         }
		        {\chi_{P_{0}}(X,t)}
	\right]
		\geq 0.
	\label{dS}
\end{align}

Whether the spatial distribution entropy $S^{X}(t)$ increases or decreases, it is completely compensated by the first term of the time derivative of $S_{I}^{X:P}(t)$, so it does not contribute to the change in the total entropy $S(t)$. 
Eq.~\eqref{dS} can be derived by a procedure similar to the proof of the H-theorem (see Appendix~\ref{appendix H-theorem}), noting that $\chi_{P_{0}}(X,t)$ follows the advection-diffusion equation. The final inequality holds because $\chi_{P_{0}}(X,t) \geq 0$ in the situation considered in Sec.~\ref{Advection-induced spatial contraction in 1D quantum Brownian motion}.

\begin{figure}[t]
\centering{}
\includegraphics[scale=0.55]{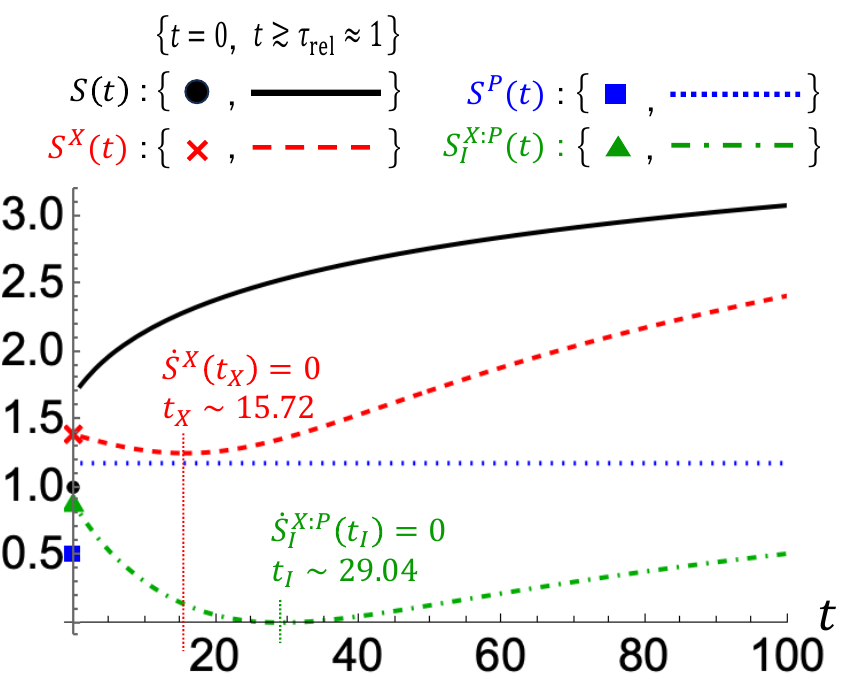}
\caption{
\label{fig_entropy_balance}
The time evolution of entropies in the situation discussed in Fig.~\ref{apparently_negative_diffusion} is shown at the initial time $t=0$ and in the region $t\gtrsim\tau_\mathrm{rel}$. 
The black circle represents $S(0)$, the solid line represents $S(t\gtrsim \tau_\mathrm{rel})$, 
the red cross represents $S^X(0)$, the dashed line represents $S^X(t\gtrsim \tau_\mathrm{rel})$, 
the blue square represents $S^P(0)$, the dotted line represents $S^P(t\gtrsim \tau_\mathrm{rel})$, 
and the green triangle represents $S_I^{X:P}(0)$, with the dot-dashed line representing $S_I^{X:P}(t\gtrsim \tau_\mathrm{rel})$.
Note that although the black solid line and blue dotted line appear to show a discontinuous jump between $t=0$ and $t=\tau_{\mathrm{rel}}$, these values are actually continuous.
This apparent discontinuity arises because the momentum relaxation process occurs on a much shorter time scale than that of spatial relaxation.
We do not plot the entropy evolution during this rapid relaxation period, as no analytical expression is available in this interval.
The time at which the entropy of the spatial distribution $S^X(t)$ reaches its minimum is $t_X\sim 15.72$ (in units of $t_{\mathrm u}$), and the time at which the mutual information $S_I^{X:P}(t)$ reaches its minimum is $t_I\sim 29.04$. 
These times coincide with $t_d$ and $t_a$ in Fig.~\ref{fig_phenomenological_diffusion_constant}.
}
\end{figure}

The entropy balance in the case discussed in Sec.~\ref{Advection-induced spatial contraction in 1D quantum Brownian motion} is shown in Fig.~\ref{fig_entropy_balance}.
Note that the figure depicts the values of each entropy term at the initial time $t=0$, as well as their changes over time in the region $t\gtrsim\tau_\mathrm{rel}$. 
The spatial distribution entropy $S^{X}(t)$ decreases for a while and then begins to increase. This behavior is consistent with the behavior of the wave packet in Fig.~\ref{apparently_negative_diffusion}, where the wave packet initially contracts and then starts expanding. 
At the same time, the mutual information $S_{I}^{X:P}(t)$ also decreases for a short time, before it begins to increase. 
As a result, the decrease in $S^{X}(t)$ is compensated, and the total entropy $S(t)$ continues to increase monotonically. 
Additionally, since the momentum distribution has relaxed to equilibrium at $t=\tau_\mathrm{rel}$, the momentum entropy $S^P(t)$ remains at its maximum value for $t\gtrsim \tau_\mathrm{rel}$ without further change.

In this entropy balance of the Brownian motion in the one-dimensional quantum system, the mutual information between the spatial distribution and momentum distribution formed in the initial state is maintained even in the local equilibrium state, and the mutual information reduces the entropy of the spatial distribution. This means the uncertainty regarding the particle's coordinate can be reduced by utilizing the correlation between coordinate and momentum. 
Once the initial correlation is established, this reduction in entropy of spatial distribution occurs naturally as the system evolves with a fixed Hamiltonian, requiring no external energy input, making the process highly efficient. 
This indicates that Brownian particles can be controlled in an energy-efficient manner with minimal heat generation, offering advantages in realizing a forward bias and the completion of the computation in Brownian computers.

This aligns with the framework of an information ratchet \cite{boyd2016identifying,sanchez2019autonomous}, where state changes are driven by information with minimal external energy supply.
In an information ratchet, the entropy of a spatial distribution can be reduced through feedback control based on the system's internal information, without directly performing work on the system. This approach is rooted in the concept of Maxwell's demon, which uses information to perform thermodynamic operations. In classical Brownian motion systems and other typical thermodynamic systems, an overdamped regime is assumed, where the momentum distribution quickly relaxes to a Maxwell distribution and becomes statistically independent of the spatial distribution. As a result, it is impossible to create and utilize a correlation between coordinate and momentum. Therefore, Maxwell's demon requires an external degree of freedom, namely the demon's memory, to create a correlation with the particle's coordinate $X$. The demon uses this correlation to control the particle's position. In this case, the reduction in system entropy is explained consistently with the second law of thermodynamics through a decrease in the mutual information between the system and the external memory \cite{sagawa2018second}.

On the other hand, in 1D quantum Brownian motion, the relaxation occurs within each subspace independently, so even after local equilibrium is established, the momentum distribution and spatial distribution remain nonseparable. In other words, $X$ and $P$ remain correlated.
Therefore, the momentum $P$, which is the dynamical variable conjugate to coordinate $X$, can be used in place of an external memory. The advantage of using momentum is that it eliminates the need to introduce the abstract concept of feedback control by an external entity, such as Maxwell's demon. 
If an appropriate correlation between $X$ and $P$ is established in the initial state, the system dynamics will naturally reduce the entropy of the spatial distribution through a reversible phase-mixing effect caused by the momentum dependence of the sound velocity $\sigma(P)$.


\section{Concluding remarks
\label{Concluding remarks}}

In this paper, we have demonstrated that Brownian motion in a 1D quantum system can exhibit unidirectional transport even in the absence of any external forces, and achieve spatial distribution contraction without violating the second law of thermodynamics. The underlying mechanism of these phenomena lies in the partitioning of the particle's momentum space into subspaces that lack momentum inversion symmetry. Such partitioning leads to the preservation of an initial correlation between coordinate and momentum even after momentum relaxation. This remaining correlation, in turn, plays a crucial role in enabling entropy reduction in the spatial distribution. Notably, the reduction is driven by a reversible phase-mixing effect that arises from the momentum dependence of the transport velocity in each momentum subspace.

These phenomena arise as a consequence of quantum dissipation in the spatially constrained 1D system. Because the collision term vanishes in the classical limit, the dissipation in the 1D system is purely a quantum effect, and these phenomena inherently require both one-dimensionality and quantum nature.

To induce spatial distribution contraction, we employed a single nonfactorizable Gaussian wave packet as the initial state. In this case, the corresponding Wigner distribution function contains no negative regions, allowing it to be regarded as a joint probability distribution in coordinate-momentum space. This makes it possible to define a conventional nonequilibrium entropy using the Wigner distribution function. Consequently, the relative entropy, which quantifies coordinate-momentum correlations as mutual information, naturally emerges in the analysis. This provides a consistent explanation for the decrease in spatial distribution entropy, without violating the second law of thermodynamics.

By the way, when discussing Brownian computers, it seems that in many cases it is implicitly understood that classical 1D Brownian motion exists based on phenomenological arguments. However, it is necessary to verify whether classical 1D Brownian motion is actually possible under the laws of microscopic physics. In fact, in the setting we have shown in this paper, the collision operator becomes zero in the classical 1D model, and classical 1D Brownian motion does not exist.

Therefore, it should be noted that the comparison of existing models of Brownian computers discussed below with our quantum model is conducted under the hypothesis that 1D Brownian motion exists in classical systems as well.

Brownian computers, which harness thermal fluctuations near equilibrium, have been expected to achieve ultra-low-energy computation. However, directing classical Brownian motion incurs certain costs. Specifically, because the Maxwellian momentum distribution is inversion-symmetric, the particle does not possess a statistical ``direction,'' making it necessary to induce drift not only to drive the computation forward but also to increase the reliability of completing computations.

By contrast, the Brownian motion in a 1D quantum system offers advantages in terms of manageability compared to its classical counterpart. The asymmetric Maxwellian distributions formed in each momentum subspace result in a nonzero transport velocity (i.e., sound velocity $\sigma(P)$). The transport coefficients depend on the momentum. Moreover, because relaxation occurs separately in each subspace, coordinate-momentum correlations present in the initial state persist even after momentum relaxation. Through appropriate preparation of the initial state, it is thus possible to control the Brownian particle near thermal equilibrium without requiring extensive external intervention.

Let us consider a comparison with scenarios resembling Maxwell's demon—so-called ``information ratchets'' \cite{boyd2016identifying,sanchez2019autonomous}—where classical Brownian particles are guided via external observation and feedback control. In such settings, an external memory is introduced to correlate the particle's coordinate with that memory, thereby controlling the Brownian particle. 
On the other hand, in our 1D quantum system, there exists an intrinsic correlation between the coordinate and its conjugate dynamical variable, momentum. As a result, the system's internal dynamics alone can effectively steer the Brownian particle without the need for an external feedback mechanism. 

Importantly, the presence of partitioning in momentum space, which is a manifestation of the 1D quantum effect, is temperature-independent. 
It is therefore possible that quantum effects remain even at relatively high temperatures, and the resulting one-way transport remains valid. 
To evaluate the practical relevance of our findings, we estimate the characteristic temperature scale $T_{\mathrm u}=mc^2/k_{\mathrm B}$, at which unidirectional transport of the 1D quantum Brownian particle remains effective (see Eqs.~\eqref{crossover temperature} and~\eqref{high-temperature condition for the particle}).

Using the parameters of an $\alpha$-helical protein molecular chain \cite{christiansen1990davydov,scott1992davydov,tanaka2009emergence}, 
the temperature can be estimated as $T_{\mathrm u}\sim 200 \ {\mathrm K}$, where the effective mass is $m \simeq 2\times 10^{-28} \ {\mathrm{kg}}$, and the propagation speed of acoustic phonons is $c \simeq 4000 \ {\mathrm{m/s}}$. In this unit, the room temperature $T=300 \ {\mathrm K}$ corresponds to $T/T_{\mathrm u}=1.5$, indicating that the unidirectional transport remains effective even at such temperatures (see Fig.~\ref{fig_sound_velocity}). When estimated using semiconductor parameters \cite{mahan1993many}, for example those of GaAs, the unit temperature becomes $T_{\mathrm u}\sim 0.09 \  {\mathrm K}$, with $m\simeq 6\times 10^{-32} \ {\mathrm{kg}}$ and $c\simeq 5000 \ {\mathrm{m/s}}$.
This value is considerably lower than that of the protein system. The difference in temperature scale is primarily due to the difference in the effective mass $m$ of the particle: in the $\alpha$-helical protein, the particle is a vibrational exciton of a C=O molecule, whereas in semiconductors, the particle is an electron with a much smaller effective mass.
However, our results also suggest that the temperature can be increased by using ions instead of electrons, even in the case of semiconductors.

As a related study, a technique called ``momentum computing'' has been proposed \cite{ray2021non,ray2023gigahertz}. This approach aims to realize computation by switching between Brownian motion and dynamically reversible (harmonic) motion, addressing the difficulties of Brownian computing arising from the lack of a statistical ``direction'' in the Brownian motion. Doing so requires temporarily decoupling the system from the thermal bath to create conditions under which the particle can move reversibly by its momentum. In contrast, in the 1D quantum Brownian motion described here, the system remains continuously coupled to the thermal bath, and unidirectional transport is achieved simply by setting the initial momentum state. Without needing to disconnect from the thermal bath, the interplay of quantum dissipative effects and the choice of initial conditions allows one to steer the Brownian particle's motion.

In this study, we have not discussed the cost associated with preparing the initial states necessary to achieve unidirectional transport or spatial distribution contraction. Evaluating such costs and considering the applicability of these conditions in experimental settings remain important challenges for future research. Nonetheless, as shown in this paper, the finding that quantum dissipation in the 1D system can spontaneously give rise to unidirectional transport without contradicting the second law of thermodynamics provides a new avenue for ultra-low-energy computation and will strongly motivate further developments in this field.

\begin{acknowledgments}
We thank Hiroaki Umehara for pointing out the potential of using the transport properties of 1D quantum Brownian motion for ratchet applications, which motivated us to pursue the analysis further.
This work was supported by JST, CREST Grant Number JPMJCR20C1, Japan.
\end{acknowledgments}

\appendix
\section{Temperature dependence of nonzero sound velocity via phonon quantum statistics
\label{appendix phonon quantum effect}}
In the main text, we demonstrated that the one-dimensional quantum nature contributes to the emergence of unidirectional transport of the Brownian particle. Specifically, in the 1D system, the particle undergoes each momentum transition through the emission or absorption of a single quantized phonon, leading to the partitioning of the particle's momentum space into disjoint subspaces that lack inversion symmetry. This partitioning is responsible for the emergence of the nonzero sound velocity.

Here, we focus on the high-temperature condition under which the quantum-statistical distribution of phonons is approximated by a classical distribution, and qualitatively show that the sound velocity approaches zero under this condition. 
For a more detailed and quantitative discussion based on the analytical solution of the sound velocity, refer to our previous work \cite{nakade2020anomalous}.
Note, however, that the connection to the quantum-statistical nature of phonons is newly addressed in this appendix.

The Bose--Einstein distribution~\eqref{n(q)} indicates that lower-energy phonon modes are predominantly occupied at low temperatures, whereas higher-energy modes become increasingly populated as the temperature increases.
In high-temperature regime, 
\begin{equation} 
   \hbar \omega_q \ll k_{\mathrm B}T,
   \label{high-temperature condition 1}
\end{equation} 
the distribution can be approximated by the classical Rayleigh--Jeans form,
\begin{equation}
   n(q) \simeq \frac{k_{\mathrm B} T}{\hbar \omega_q},
   \label{classical Rayleigh--Jeans form}
\end{equation}
where virtually all phonon modes are thermally excited, and each mode carries an average energy of $k_{\mathrm B} T$.

\begin{figure}[t]
\centering{}
\includegraphics[scale=0.6]{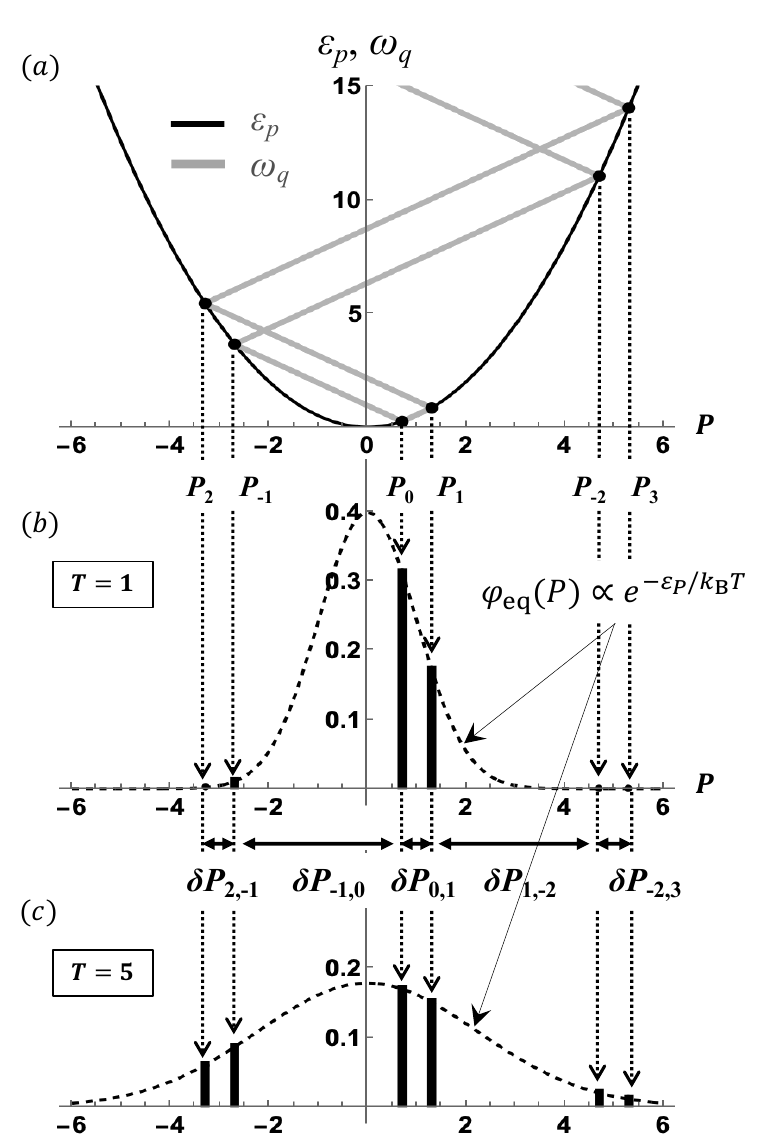}
\caption{
\label{fig_1D_quantum_transitions}
Discrete momentum transitions in the 1D quantum system and the resulting asymmetric Maxwell distribution. All units are as described in the model introduction. (a) Discrete momentum subset accessible from $P_0=0.7$. (b) Momentum equilibrium distribution in the subspace ${\cal S}_{P_0=0.7}$ at $T=1$ in the temperature unit $T_{\mathrm u}$. Only a few momentum states near the origin are thermally excited. (c) Momentum equilibrium distribution in the same subspace at $T=5$. A broader range of momentum states is excited.
}
\end{figure}

The high-temperature condition~\eqref{high-temperature condition 1} can be rewritten by substituting the phonon dispersion relation~\eqref{dispersion relation} as
\begin{equation}
   c\hbar |q| \ll k_{\mathrm B}T.
   \label{high-temperature condition 2}
\end{equation}

As explained in Sec.~2.2, the emission or absorption of a single phonon (carrying momentum $\hbar q$) leads to discrete momentum transitions of the particle in the 1D system. Due to momentum conservation and the resonance condition, the accessible discrete momenta $P_{\nu}$ form a subset~\eqref{momentum subset}.
 
In this framework, a single phonon-induced momentum transition is given by
\begin{equation}
   \hbar |q| = \bigl | P_{\nu\pm1} - P_\nu \bigr |.
\end{equation}
Substituting this into the high-temperature condition~\eqref{high-temperature condition 2} and rearranging in terms of the momentum unit $p_{\mathrm u}=mc$ and temperature unit $T_{\mathrm u}=mc^2/k_{\mathrm B}$, we obtain
\begin{equation}
   \frac{\bigl | P_{\nu \pm 1} - P_\nu \bigr |}{p_{\mathrm u}}  \ll  \frac{T}{T_{\mathrm u}}.
      \label{high-temperature condition 3}
\end{equation}
Under this condition, the phonon distribution approaches the classical form~\eqref{classical Rayleigh--Jeans form}, and phonons with larger momentum $\hbar |q|$ are thermally excited. As a result, the particle can undergo larger momentum transitions via phonon absorption and emission.

This behavior can be understood by examining the equilibrium momentum distribution of the particle, as shown in Fig.~\ref{fig_1D_quantum_transitions}.
As the temperature increases, the width of the Maxwellian momentum distribution expands, increasing the probability of transitions between states with larger momentum difference:
\begin{equation}
   \Delta P_{\nu \pm 1,\nu}:=|P_{\nu \pm 1} - P_\nu|.
\end{equation}
Consequently, a greater number of discrete momentum states are thermally excited.
Note that the momentum difference $\Delta P_{\nu \pm 1,\nu}$ corresponds to the transition between the dots connected by gray lines in Fig.~\ref{fig_1D_quantum_transitions}(a), and does not represent the transition between adjacent discrete momentum states.

At a given temperature, the typical upper bound of momentum difference, $\Delta P_{\mathrm{upper}}$, can be roughly estimated by the width of the Maxwellian distribution:
\begin{equation}
  \Delta P_{\mathrm{upper}} \sim 2\sqrt{mk_{\mathrm B}T}
\end{equation}
Furthermore, the interval between two adjacent discrete momenta within a subspace can be expressed as
\begin{equation}
   \delta P_{\nu,-(\nu \pm 1)} :=\bigl | P_{\nu} - P_{-(\nu \pm 1)} \bigr | = 2\bigl | P_0 - mc\bigr |,
\end{equation}
where $P_0$ is the momentum satisfying $-mc \leq P_0 \leq mc$, leading to the following range:
\begin{equation}
   0 \leq 2\bigl | P_0 - mc\bigr |  \leq 4mc.
\end{equation}
Thus, the average interval is typically of the order of $\overline{\delta P} \sim 2mc$.

The number of discrete momentum states that are thermally excited at a given temperature can be estimated as follows:
\begin{equation}
   \frac{\Delta P_{\mathrm{upper}}}{\overline{\delta P}}
   \sim
   \frac{2\sqrt{mk_{\mathrm B}T}}{2mc}=\sqrt{\frac{T}{T_{\mathrm u}}}.
\end{equation}
Accordingly, the system can be regarded as ``low-temperature'' if this number is of order unity,
\begin{equation}
   \sqrt{\frac{T}{T_{\mathrm u}}} \sim 1,
\end{equation}
and ``high-temperature'' if this number is much greater than unity,
\begin{equation}
   \sqrt{\frac{T}{T_{\mathrm u}}} \gg 1.
   \label{high-temperature condition for the particle}
\end{equation}

This high-temperature condition, derived from the particle's momentum distribution, corresponds to the phonon-based high-temperature condition~\eqref{high-temperature condition 3} when the representative momentum transition is set to
\begin{equation}
   |P_{\nu \pm 1} - P_\nu| \sim mc.
\end{equation}
Here, $mc$ corresponds to half of the average interval $\overline{\delta P}\sim 2mc$ between adjacent discrete momentum states. 
When the temperature is on the order of $mc$, only one or two discrete momentum states that are very close together (i.e., with the same sign) are thermally excited, as shown in Fig.~\ref{fig_1D_quantum_transitions} (b).
Consequently, the contributions from positive and negative momenta do not cancel out, and the sound velocity remains nonzero. 
Therefore, both conditions are qualitatively consistent: increasing the temperature leads to the thermal excitation of more phonon modes with larger momentum, thereby allowing the particle to access a broader range of momentum states.

\begin{figure}[t]
\centering{}
\includegraphics[scale=0.6]{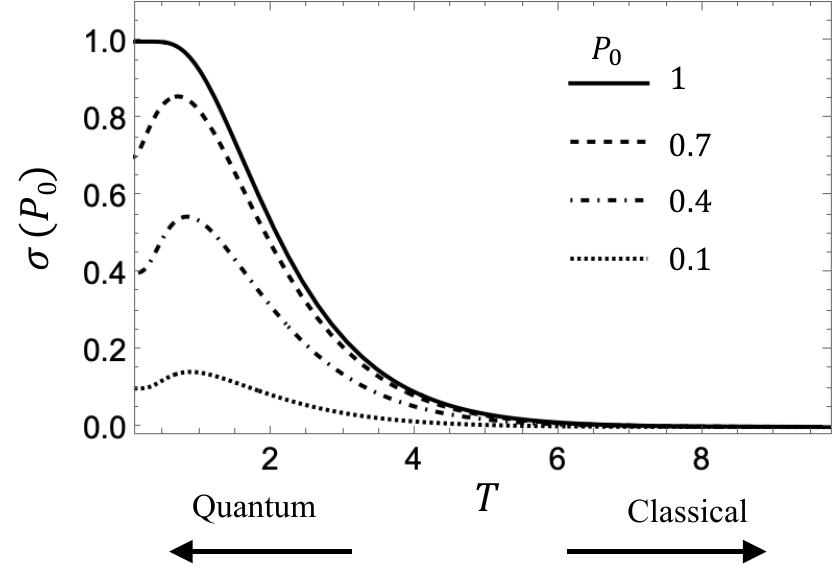}
\caption{
\label{fig_sound_velocity}
Temperature dependence of the sound velocity $\sigma(P_0)$. The units of temperature and momentum are $T_{\mathrm u}=mc^2/k_{\mathrm B}$, and $p_{\mathrm u}=mc$, respectively. The sound velocity is defined within each momentum subspace ${\cal S}_{P_0}$. The solid line corresponds to $P_0=1$, the dashed line to $P_0=0.7$, the dashed-dotted line to $P_0=0.4$, and the dotted line to $P_0=0.1$. The sound velocity takes large value around $T\sim 1$, and asymptotically vanishes in high-temperature regime $T\gg1$.
}
\end{figure}

Now, let us focus on the temperature dependence of the sound velocity~\eqref{sound velocity 2}.
The sound velocity is defined as the average momentum in units in which $m=1$, weighted by the Maxwellian momentum distribution over the discrete momentum states within the subspace ${\cal S}_{P_0}$.

At low temperatures, only a few discrete momentum states near the origin are significantly populated. Due to the inversion asymmetry of the subspace ${\cal S}_{P_0}$, their contributions to the sound velocity do not cancel out, resulting in a nonzero value.
As the temperature increases, however, a broader range of momentum states becomes thermally excited.
As a result, the positive and negative momentum contributions become more balanced, leading to cancellation in the average, and the sound velocity gradually decreases toward zero.
This behavior is illustrated in Fig.~\ref{fig_sound_velocity}.

In the high-temperature regime~\eqref{high-temperature condition for the particle}, i.e., $T/T_{\mathrm u} \gg 1$, the sound velocity asymptotically approaches zero, as shown in Fig.~\ref{fig_sound_velocity}. Thus, in this regime, the particle behaves similarly to the phenomenologically hypothesized 1D classical Brownian motion, in which no unidirectional transport occurs.

In summary, the characteristic temperature $T_{\mathrm u}$ defined as a unit temperature characterizes the quantum-to-classical crossover in this system.

\section{Threshold temperature for the spatial contraction
\label{appendix threshold temperature}}

In this appendix, we identify the threshold temperature below which the system, with the initial distribution~\eqref{initial squeezed Gaussian}, exhibits spatial contraction.

For this analysis, we adopt parameter values relevant to the $\alpha$-helical protein molecular chain~\cite{christiansen1990davydov,scott1992davydov}.
This choice is motivated by the fact that, in the protein molecular chain, the ratchet effect remains observable at temperatures on the order of $10^2 {\rm K}$, while for semiconductor parameters, it becomes significant only at much lower temperatures on the order of $10^{-1} {\rm K}$.

In $\alpha$-helical protein molecular chain system, the characteristic scale of the diffusion coefficient~\eqref{D_u} is estimated as $D^{\ast}\sim 0.01 {\mathrm{cm^2/s}}$, with the deformation potential $\Delta_0 \simeq 0.3 \ {\mathrm{eV}}$ and the 1D mass density $\rho_M \simeq 1\times 10^{-15} \ {\mathrm{kg/m}}$. Note that under these parameters, the ratio between the characteristic diffusion scale and the diffusion unit is approximately $D^{\ast}/D_{\mathrm u} \sim 2$.

As discussed in Sec.~\ref{Advection-induced spatial contraction in 1D quantum Brownian motion}, the spatial contraction can be observed during the time interval~\eqref{time interval}.
For such a time interval to exist, the inequality $\tau_\mathrm{rel}< t_d$ must be satisfied.
Using Eq.~\eqref{td}, this inequality can be rewritten as
\begin{align}
   D^{(x)}(\tau_{\mathrm{rel}})<0.
\end{align}

\begin{figure}[htbp]
\centering{}
\includegraphics[scale=0.5]{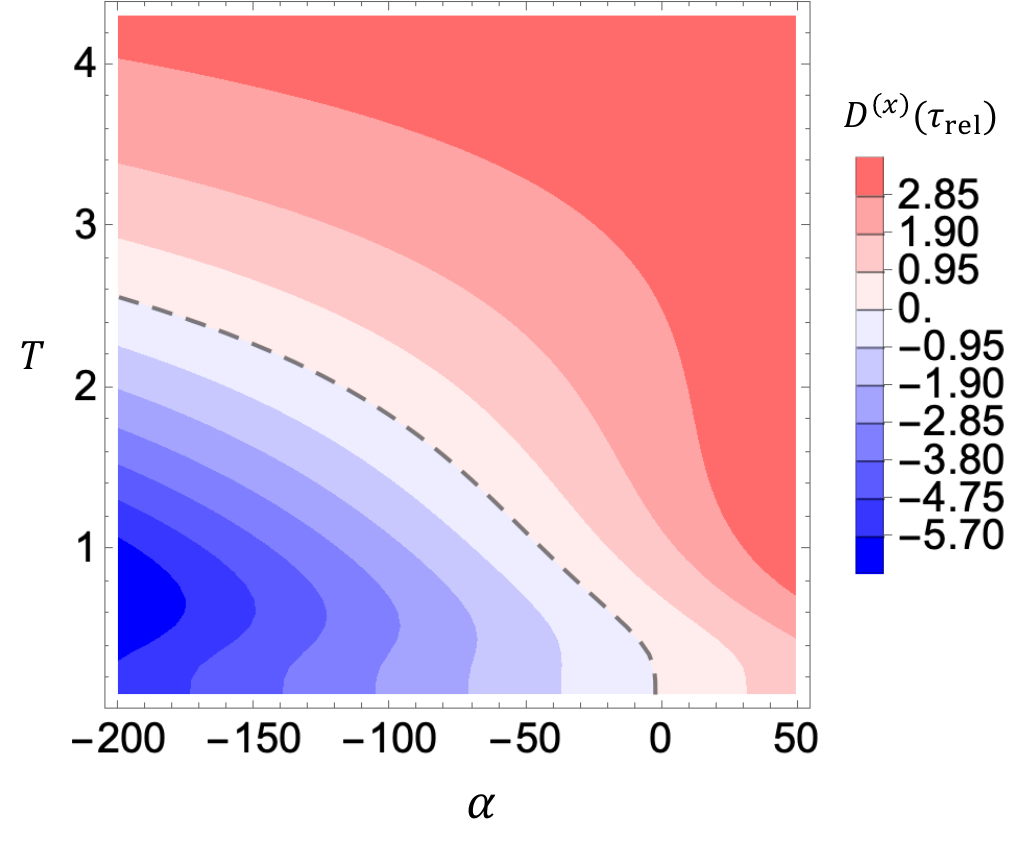}
\caption{
\label{fig_parameter_dependence_of_spatial_contraction}
Contour plot of $D^{(x)}(\tau_{\mathrm{rel}})$ for an initial nonfactorizable Gaussian wave packet~\eqref{initial squeezed Gaussian}, centered at $(X^{\prime}, P^{\prime})=(0,0)$ with width $\widetilde{\Delta X}=3$. The dashed line shows the threshold at which the value becomes zero. This plot shows the $\alpha$--$T$ parameter region where the spatial contraction occurs. 
The values are computed using a parameter set representative of a protein molecular chain. In this setting, $D^{\ast}/D_{\mathrm u}\sim 2$.
All quantities are expressed in units introduced in Sec.~\ref{Model and Kinetic equation}: the unit of the diffusion coefficient is $D_{\mathrm u}=x_{\mathrm u}^2/t_{\mathrm u}\sim 6\times10^{-3} {\mathrm{cm^2/s}}$, the unit of temperature is $T_{\mathrm u}\sim 200\ {\mathrm K}$, and the unit of the initial shift parameter $\alpha$ is $x_{\mathrm u}/p_{\mathrm u}=\hbar/m^2c^2\sim 2\times10^{14} \ {\mathrm{s/kg}}$, with the effective mass $m \simeq 2\times 10^{-28} \ {\mathrm{kg}}$, and the acoustic phonon propagation speed $c \simeq 4000 \ {\mathrm{m/s}}$. The large unit value of $\alpha$ is due to the small effective mass of the molecular system.}
\end{figure}

By varying the temperature $T$ and initial shift parameter $\alpha$ and plotting $D^{(x)}(\tau_{\mathrm{rel}})$, we obtain a contour plot as shown in Fig.~\ref{fig_parameter_dependence_of_spatial_contraction}.
The bottom-left (blue) region represents the parameter space where $D^{(x)}(\tau_{\mathrm{rel}})<0$, indicating that the advection effect dominates over diffusion and the system exhibits spatial contraction. The dashed contour line corresponds to the threshold where $D^{(x)}(\tau_{\mathrm{rel}})=0$.
In the top-right (red) region above this line, the diffusion dominates and the system does not exhibit spatial contraction.

As estimated in the concluding remarks, room temperature, $300 \ {\mathrm K}$, corresponds to $T/T_{\mathrm u}= 1.5$ under the protein parameters.
Even at such relatively high temperatures, the advection effect can still dominate over the diffusion effect, and spatial contraction can occur.

%

\section{H-theorem \label{appendix H-theorem}}

We prove that the H-theorem holds
when the Wigner distribution function follows the advection-diffusion Eq.~\eqref{advection-diffusion} with the condition~\eqref{D(P_0)>0}.

It should be noted that, in general, the Wigner distribution function $f^W$ is a quasi-probability distribution that can take negative values  \cite{hudson1974wigner}. 
Consequently, the functional defined as $-\int \!dXdP\  f^W\ln f^W$, which is conventionally used as the H-function in classical system,
has an imaginary part.
Therefore, it is necessary to identify another functional that 
is real and  
monotonically increases as the H-function.

In this proof, we assume that
the Wigner distribution function satisfies the boundary conditions:
\begin{align}
	\lim_{X\rightarrow\pm\infty}f^{W}(X,P,t)&=0,
	\label{boundary condition 1}\\
	\lim_{X\rightarrow\pm\infty}\frac{\partial}{\partial X}f^{W}(X,P,t)&=0.
	\label{boundary condition 2}
\end{align}

We then introduce a new function which is non-negative at any $X$ and $P$
as follows:
\begin{equation}
	\tilde{f}^{W}(X,P,t):= f^{W}(X,P,t)+C>0,
	\label{bottom-up function}
\end{equation}
where the constant satisfies
\begin{equation}
	C> \frac{1}{\pi \hbar}. 
	\label{constant C}
\end{equation}
%
Since the Wigner distribution function has upper and lower bounds  \cite{leonhardt1995measuring}, as shown in Eq.~\eqref{below and upper bound of Wigner function}, the inequality in Eq.~\eqref{bottom-up function} holds.
Note that the function~\eqref{bottom-up function} also obeys the advection-diffusion Eq.~\eqref{advection-diffusion}.

Let us introduce a functional with the new function~\eqref{bottom-up function} and the Wigner distribution function as
\begin{equation}
	H_{[\tilde{f}^{W}]}(t):=
	\!
	 \int_{-\infty}^{\infty}\!\!\!\!\!\!\!dX
	 \!
	 \int_{-\infty}^{\infty}\!\!\!\!\!\!\!dP\
	  f^{W}(X,P,t)
		\ln
		\{
		\tilde{f}^{W}(X,P,t)/\frac{1}{\pi\hbar}
		\}.
	\label{H function for complete H theorem}
\end{equation}
Here, the factor $1/\pi\hbar$ inside the logarithm is for non-dimensionalization.
Note that we use $\tilde{f}^{W}(X,P,t)$ only for the argument of the natural logarithm. This is because if we replace $f^{W}(X,P,t)$ in Eq.~\eqref{H function for complete H theorem} with $\tilde{f}^{W}(X,P,t)$, then the functional diverges
because of the factor $C$
.

Since the non-dimensionalization factor $1/\pi\hbar$ does not contribute to the time evolution of $H_{[\tilde{f}^{W}]}(t)$,
we have
\begin{align}
	 &\frac{d}{dt}H_{[\tilde{f}^{W}]}(t) \nonumber \\
	 &=
	 \frac{d}{dt}
	  \int_{-\infty}^{\infty}\!\!\!\!\!\!\!dX
		\int_{-\infty}^{\infty}\!\!\!\!\!\!\!dP\
		 \left\{ \tilde{f}^{W}(X,P,t)-C\right\}
		 \ln\tilde{f}^{W}(X,P,t)
		\nonumber\\
	\nonumber\\
	&=
	 \int_{-\infty}^{\infty}\!\!\!\!\!\!\!dX
	 \int_{-\infty}^{\infty}\!\!\!\!\!\!\!dP\
	  \left\{ \frac{\partial}{\partial t}\tilde{f}^{W}(X,P,t)\right\}
		\ln\tilde{f}^{W}(X,P,t) \nonumber \\
	 &\ \ \ \ \ \ -C\int_{-\infty}^{\infty}\!\!\!\!\!\!\!dX
	   \int_{-\infty}^{\infty}\!\!\!\!\!\!\!dP\
	    \frac{\frac{\partial}{\partial t}\tilde{f}^{W}(X,P,t)}
		       {\tilde{f}^{W}(X,P,t)}
		\nonumber\\
	\nonumber\\
	&=
	 -\int_{-\infty}^{\infty}\!\!\!\!\!\!\!dP\ D(P)
	 \int_{-\infty}^{\infty}\!\!\!\!\!\!\!dX
	  \Biggl[
		 \frac{\{\frac{\partial}{\partial X}\tilde{f}^{W}(X,P,t)\}^{2}}
		      {\tilde{f}^{W}(X,P,t)}
	\nonumber\\
		 &\ \ \ \ \ \ \ \ \ \ \ \ \ \ \ \ \ \ \ \ \ \ \ \ \ \ \ \ \ \ \ 
		 +C\left\{ \frac{\frac{\partial}{\partial X}\tilde{f}^{W}(X,P,t)}
		                {\tilde{f}^{W}(X,P,t)}
			 \right\} ^{2}
		\Biggr]
	\leq 0. 
	\label{d/dt H}
\end{align}
In the transition from the second to the third line, we substituted Eq.~\eqref{advection-diffusion} and performed integration by parts over $X$ with the boundary conditions~\eqref{boundary condition 1} and~\eqref{boundary condition 2}.
It is clear that
the inequality in Eq.~\eqref{d/dt H} holds because of the
conditions~\eqref{D(P_0)>0} and~\eqref{bottom-up function}.
Hence, this functional satisfies the H-theorem.

In the time derivative of the H function~\eqref{d/dt H},
the contribution of the advection term with $\sigma(P)$ in Eq.~\eqref{advection-diffusion} disappears
and only the contribution of the diffusion term with $D(P)$ remains.
In other words, only the diffusion term in Eq.~\eqref{advection-diffusion} is essential for the decrease in the H function.

When we consider the time evolution of the Wigner distribution function of a free particle which has no interaction with the phonons,
there is no diffusion term in the transport equation.
Therefore, the time derivative of the H function~\eqref{H function for complete H theorem} is zero.
This is consistent with the fact that free particle propagation is a reversible process and there is no production of entropy.

From the above discussion, it becomes obvious that if the Wigner distribution function obeys the advection-diffusion Eq.~\eqref{advection-diffusion}, the H-theorem holds.
That is to say, there is no entropy reduction in the particle propagation at the local equilibrium in this 1D quantum system.
It should be emphasized that it is essential for the H-theorem that the diffusion coefficient $D(P)$ derived from the microscopic theory is always positive.

\section{Analytical solution of the phenomenological diffusion coefficient
\label{appendix D^x(t)}}

In this appendix, we derive the analytical solution~\eqref{analytic solution of D^x(t)} for the phenomenological diffusion coefficient~\eqref{phenomenological diffusion coefficient}, when the Wigner distribution function follows the advection-diffusion equation~\eqref{advection-diffusion}. 

The solution~\eqref{analytic solution of D^x(t)} was first derived in our previous work  \cite{nakade2020anomalous} by expressing the initial Wigner distribution function as a linear combination of Gaussians, using the theorem that any square-integrable function can be expanded in terms of Gaussians  \cite{calcaterra2008approximating}. 

In this appendix, we introduce an alternative method to obtain the same solution by calculating the $n$-th moment $\langle X^n\rangle_{t}$ using the formal solution of the Fourier components of the advection-diffusion equation~\eqref{advection-diffusion}. 
While the form of the resulting solution is the same, the derivation in this appendix clarifies why the second term in $D^{(x)}(t\gtrsim\tau_\mathrm{rel})$ depends on the initial distribution.
The reason is due to momentum relaxation occurring independently within each subspace ${\mathcal S}_{P_0}$. 

The derivation consists of two stages.

In the first stage, using the relationship between the $n$-th moment with respect to the coordinate $X$ and the Fourier component, we show that the phenomenological diffusion coefficient can be expressed as the sum of two time-independent terms and one term that is linear in time. 
Each of these terms is written as the average value over the Wigner distribution function immediately after reaching local equilibrium $f^W(X,P,t=\tau_\mathrm{rel})$.

In the second stage, we analyze the relationship between the average values over $f^W(X,P,t=\tau_\mathrm{rel})$ and the average values over the initial Wigner distribution function $f^W(X,P,t=0)$, deriving the analytical solution~\eqref{analytic solution of D^x(t)}.

\subsection{First stage: Transforming $D^{(x)}(t)$ in terms of the Fourier component}
We transform the phenomenological diffusion coefficient,
\begin{equation}
   D^{(x)}(t)=
   \frac{1}{2}\frac{d}{dt}
   (\langle X^{2}\rangle_{t}-\langle X\rangle_{t}^{2}),
   \label{phenomenological diffusion coefficient 1}
\end{equation}
by using the relationship between the $n$-th moment with respect to the coordinate $X$ and the Fourier component: 
\begin{align}
    \langle X^n\rangle_{t}
     :=
     &
                             \int_{-\infty}^{\infty}\!\!\!\!\!\!\!dP
                             \int_{-\infty}^{\infty}\!\!\!\!\!\!\!dX\ 
                                  X^n f^{W}(X,P,t)
                              \nonumber\\
      =
      &\ 
                             i^n\int_{-\infty}^{\infty}\!\!\!\!\!\!\!dP\ 
                             \left.
                                    \frac{\partial^n}{\partial k^n}
                                    f_{k}(P,t)
                             \right|_{k=0},
\label{eq:average of X^n}
\end{align}
where $n=0,1,2,\cdots$. 

Rewriting the advection-diffusion equation~\eqref{advection-diffusion} for its corresponding Fourier component gives
\begin{equation}
    \frac{\partial}{\partial t} 
    f_{k}(P,t) 
    =\left\{
             -ik\sigma(P)-k^{2}D(P)
      \right\} 
       f_{k}(P,t),
\end{equation}
and the formal solution for the equation is
\begin{equation}
    f_{k}(P,t\geq\tau_\mathrm{rel})
    =e^{\{-ik\sigma(P)-k^{2}D(P)\}(t-\tau_\mathrm{rel})}
    f_{k}(P,t=\tau_\mathrm{rel}).
    \label{solution of the fourie component of the Wigner distribution function}
\end{equation}
Note that the formal solution applies for $t\geq\tau_\mathrm{rel}$, i.e., after local equilibrium is reached.

We calculate the first and second moments with respect to the coordinate $X$ by
substituting~\eqref{solution of the fourie component of the Wigner distribution function} into~\eqref{eq:average of X^n}:
\begin{align}
    \langle X\rangle_{t\geq\tau_\mathrm{rel}} 
    =&\ 
     \langle X\rangle_{t=\tau_{\mathrm{rel}}}
     +(t-\tau_{\mathrm{rel}})
            \langle\sigma(P)\rangle_{t=\tau_\mathrm{rel}},
\\
    \langle X^{2}\rangle_{t\geq\tau_\mathrm{rel}} 
    =&\ 
    \langle X^{2}\rangle_{t=\tau_{\mathrm{rel}}}
\nonumber\\
    &+2(t-\tau_{\mathrm{rel}})
         \left\{
          \langle D(P)\rangle_{t=\tau_{\mathrm{rel}}}
          +
          \langle X\sigma(P)\rangle_{t=\tau_{\mathrm{rel}}} 
         \right\}
\nonumber\\
    &
  +(t-\tau_{\mathrm{rel}})^{2}
          \langle(\sigma(P))^{2}\rangle_{t=\tau_{\mathrm{rel}}},
\end{align}
where $\langle\cdot\rangle_{t=\tau_\mathrm{rel}}$ indicates the average over the Wigner distribution function at $t=\tau_\mathrm{rel}$.

The phenomenological diffusion coefficient~\eqref{phenomenological diffusion coefficient 1} can be rewritten as
\begin{align}
    D^{(x)}(t & \geq \tau_{\mathrm{rel}})
    =
    \nonumber\\
    & \langle D(P)\rangle_{t=\tau_{\mathrm{rel}}}
    \nonumber\\
    & +\left\langle (
                           X-\langle X\rangle_{t=\tau_{\mathrm{rel}}}
                           )
                          (
                          \sigma(P)-\langle\sigma(P)\rangle_{t=\tau_\mathrm{rel}}
                           )
        \right\rangle _{t=\tau_{\mathrm{rel}}}\nonumber
  \\&
      +(t-\tau_{\mathrm{rel}})
          \bigl\langle 
                          (
                          \sigma(P)-\langle\sigma(P)\rangle_{t=\tau_\mathrm{rel}}
                          )^{2}
           \bigr\rangle_{t=\tau_\mathrm{rel}}.
    \label{phenomenological diffusion coefficient with tau_rel average}
\end{align}
Here, each term represents the average over the Wigner distribution function at $t=\tau_\mathrm{rel}$.

\subsection{Second stage: Initial distribution dependence of each term
\label{Second stage: Initial distribution dependence of each term}}

Firstly, we show that the averages of the transport coefficients are determined by the initial distribution and are conserved at any time as follows:
\begin{align}
    \langle D(P)\rangle_{t>\tau_\mathrm{rel}} & 
    =\langle D(P)\rangle_{t=\tau_\mathrm{rel}}
    =\langle D(P)\rangle_{t=0}
    =\bar{D}
    ={\mathrm{const.}},\nonumber \\
    \langle \sigma(P)\rangle_{t>\tau_\mathrm{rel}} & 
    =\langle \sigma(P)\rangle_{t=\tau_\mathrm{rel}}
    =\langle \sigma(P)\rangle_{t=0}
    =\bar{\sigma}
    ={\mathrm{const.}}.
    \label{eq: averages of the transport coefficients}
\end{align}

The first equality, $\langle \cdot \rangle_{t>\tau_\mathrm{rel}}=\langle \cdot\rangle_{t=\tau_\mathrm{rel}}$, holds 
because the momentum distribution does not change after the local equilibrium is reached [See Eq.~\eqref{phi_P}].
To prove the second equality, we use the fact that in the 1D quantum system,
any discrete momentum $P_\nu$ within the subspace ${\mathcal S}_{P_0}$ shares the same transport coefficients [See Eq.~\eqref{momentum dependency of the transport coefficients}].
\begin{align}
 \langle D&(P)\rangle_{t=\tau_{\mathrm{rel}}}
 \nonumber\\
 & =\int_{-\infty}^{\infty}\!\!\!\!\!\!\!dX
      \int_{-mc}^{mc}\!\!\!\!\!\!\!dP_{0}\sum_{\nu=-\infty}^{\infty}\ 
             D(P_{\nu})f^{W}(X,P_{\nu},t=\tau_{\mathrm{rel}})\nonumber\\
 & =\int_{-mc}^{mc}\!\!\!\!\!\!\!dP_{0}\ 
             D(P_0)
       \int_{-\infty}^{\infty}\!\!\!\!\!\!\!dX\ 
             \chi_{P_{0}}(X,t)
    \sum_{\nu=-\infty}^{\infty} 
    \varphi_{P_{0}}^{ {\mathrm{eq}}}(P_{\nu})\nonumber\\
  & =\int_{-mc}^{mc}\!\!\!\!\!\!\!dP_{0}\ 
             D(P_{0})
      \sum_{\mu=-\infty}^{\infty}
             \varphi(P_{\mu}(P_0),0)
    \nonumber\\
  & =\int_{-\infty}^{\infty}\!\!\!\!\!\!\!dX
       \int_{-mc}^{mc}\!\!\!\!\!\!\!dP_{0}\sum_{\mu=-\infty}^{\infty}\ 
             D(P_{\mu})f^{W}(X,P_{\mu}(P_0),t=0)
\nonumber\\
  & =\langle D(P)\rangle_{t=0}.
    \label{conservation of averages of the transport coefficients}
\end{align}
In the transition from the second to the third line, we use Eq.~\eqref{varphiinit}.
The proof of Eq.~\eqref{varphiinit} is provided below.
\begin{align}
  &\int \! dX\ 
    \chi_{P_{0}}(X,t)
 \nonumber\\
  &=
  \int \! dX\ 
  \frac{1}{2\pi}
    \int_{-\infty}^{\infty}\!dk\ 
      e^{ikX}
      e^{
         -\{ik\sigma(P_{0})+k^{2}D(P_{0})\}t
         }
 \nonumber\\   
  &\ \ \ \ \ \ \ \ \ \ \ \ \ \ \ \ \ \ \ \ \ \ \ \ \ \ \ \ \ 
  \times
  \sum_{\mu=-\infty}^{\infty}
        f_{k}(P_{\mu}(P_0),0)\nonumber\\
  &=
    \int_{-\infty}^{\infty}\!dk\ 
      \delta(k)
      e^{
         -\{ik\sigma(P_{0})+k^{2}D(P_{0})\}t
         }
  \sum_{\mu=-\infty}^{\infty}
        f_{k}(P_{\mu}(P_0),0)\nonumber\\
  &=
  \sum_{\mu=-\infty}^{\infty}
        f_0(P_{\mu}(P_0),0)
=
  \sum_{\mu=-\infty}^{\infty}
    \varphi(P_{\mu}(P_0),0).
    \label{proof of varphiinit}
\end{align}

Secondly, we show that the second term in Eq.~\eqref{phenomenological diffusion coefficient with tau_rel average} can be decomposed as follows:
\begin{align}
   &\left\langle (
                           X-\langle X\rangle_{t=\tau_{\mathrm{rel}}}
                           )
                          (
                          \sigma(P)-\bar{\sigma}
                           )
        \right\rangle_{t=\tau_{\mathrm{rel}}}
   \nonumber\\
   &\
   \ \ \ \ \ \ \ \ \ \ = 
   \left\langle 
     (X-\langle X\rangle_{t=0})
     (
     \sigma(P)- 
     \bar{\sigma}
     )
   \right\rangle _{t=0}
\nonumber\\
   &\ 
   \ \ \ \ \ \ \ \ \ \ \ \ \ +
   \tau_\mathrm{rel}\ 
     \bigl\langle 
       (
       \sigma(P)-
        \bar{\sigma}
       )^{2}
     \bigr\rangle_{t=\tau_{\mathrm{rel}}}.
\label{decomposition of the second term}
\end{align}

The left-hand side of the above equation can be decomposed as
\begin{align}
   \left\langle
                      X\sigma(P)
   \right\rangle _{t=\tau_{\mathrm{rel}}}
   -
   \left\langle
                      X
   \right\rangle _{t=\tau_{\mathrm{rel}}}
   \bar{\sigma}.
\label{transformation of the second term}
\end{align}
By using the relation~\eqref{eq:average of X^n} and the Fourier component immediately after reaching local equilibrium:
\begin{align}
    &f_k(P_\nu,t=\tau_\mathrm{rel})
    \nonumber\\
    &=
     e^{-\{ik\sigma(P_{0})+k^{2} D(P_{0})\}\tau_\mathrm{rel}}
           \varphi_{P_{0}}^{ {\mathrm{eq}}}(P_{\nu}) \sum_{\mu=-\infty}^{\infty} f_{k} (P_{\mu},t=0),
\end{align}
we can transform the averages in Eq.~\eqref{transformation of the second term} as,
\begin{align}
    \langle X&\rangle_{t=\tau_\mathrm{rel}}
    \nonumber\\
    &
    =\int_{-\infty}^{\infty}\!\!\!\!\!\!\!dX
      \int_{-mc}^{mc}\!\!\!\!\!\!\!dP_{0}
      \sum_{\nu=-\infty}^{\infty}Xf^{W}(X,P_{\nu},t=\tau_{\mathrm{rel}})
    \nonumber
    \\&
    =\int_{-mc}^{mc}\!\!\!\!\!\!\!dP_{0}
      \sum_{\nu=-\infty}^{\infty}i\left.\left\{ \frac{\partial}{\partial k}f_{k}(P_{\nu},t=\tau_{\mathrm{rel}})\right\} \right|_{k=0}
    \nonumber
    \\&
    =\int_{-mc}^{mc}\!\!\!\!\!\!\!dP_{0}
      \sum_{\mu=-\infty}^{\infty}
      \{ 
      \tau_{{\mathrm{rel}}}
        \sigma(P_{\mu})
        f_{0}(P_{\mu},t=0)
\nonumber\\
       &
       \ \ \ \ \ \ \ \ \ \ \ \ \ \ \ \ \ \ \ \ \ \ +
        \left.i\frac{\partial}{\partial k}f_{k}(P_{\mu},t=0)\right|_{k=0}
        \}
    \nonumber
    \\&
    =\tau_{\mathrm{rel}}\langle\sigma(P)\rangle_{t=0}
      +\langle X\rangle_{t=0}.
      \label{decomposition of X}
\end{align}
Similarly, we can get
\begin{align}
  \langle X\sigma(P)\rangle_{t=\tau_{\mathrm{rel}}}
  =\tau_{{\mathrm{rel}}}\langle(\sigma(P))^{2}\rangle_{t=\tau_{{\mathrm{rel}}}}
    +\langle X\sigma(P)\rangle_{t=0}.
    \label{decomposition of sigmaX}
\end{align}
By substituting Eqs.~\eqref{decomposition of X} and~\eqref{decomposition of sigmaX} into Eq.~\eqref{transformation of the second term}, we obtain the right-hand side of Eq.~\eqref{decomposition of the second term}.

Finally, we get the analytical solution~\eqref{analytic solution of D^x(t)} by substituting Eq.~\eqref{decomposition of the second term} into Eq.~\eqref{phenomenological diffusion coefficient with tau_rel average}.


\begin{thebibliography}{47}%
\makeatletter
\providecommand \@ifxundefined [1]{%
 \@ifx{#1\undefined}
}%
\providecommand \@ifnum [1]{%
 \ifnum #1\expandafter \@firstoftwo
 \else \expandafter \@secondoftwo
 \fi
}%
\providecommand \@ifx [1]{%
 \ifx #1\expandafter \@firstoftwo
 \else \expandafter \@secondoftwo
 \fi
}%
\providecommand \natexlab [1]{#1}%
\providecommand \enquote  [1]{``#1''}%
\providecommand \bibnamefont  [1]{#1}%
\providecommand \bibfnamefont [1]{#1}%
\providecommand \citenamefont [1]{#1}%
\providecommand \href@noop [0]{\@secondoftwo}%
\providecommand \href [0]{\begingroup \@sanitize@url \@href}%
\providecommand \@href[1]{\@@startlink{#1}\@@href}%
\providecommand \@@href[1]{\endgroup#1\@@endlink}%
\providecommand \@sanitize@url [0]{\catcode `\\12\catcode `\$12\catcode
  `\&12\catcode `\#12\catcode `\^12\catcode `\_12\catcode `\%12\relax}%
\providecommand \@@startlink[1]{}%
\providecommand \@@endlink[0]{}%
\providecommand \url  [0]{\begingroup\@sanitize@url \@url }%
\providecommand \@url [1]{\endgroup\@href {#1}{\urlprefix }}%
\providecommand \urlprefix  [0]{URL }%
\providecommand \Eprint [0]{\href }%
\providecommand \doibase [0]{https://doi.org/}%
\providecommand \selectlanguage [0]{\@gobble}%
\providecommand \bibinfo  [0]{\@secondoftwo}%
\providecommand \bibfield  [0]{\@secondoftwo}%
\providecommand \translation [1]{[#1]}%
\providecommand \BibitemOpen [0]{}%
\providecommand \bibitemStop [0]{}%
\providecommand \bibitemNoStop [0]{.\EOS\space}%
\providecommand \EOS [0]{\spacefactor3000\relax}%
\providecommand \BibitemShut  [1]{\csname bibitem#1\endcsname}%
\let\auto@bib@innerbib\@empty
\bibitem [{\citenamefont {Bennett}(1982)}]{bennett1982thermodynamics}
  \BibitemOpen
  \bibinfo {author} {\bibfnamefont {C.~H.}\ \bibnamefont {Bennett}}\bibfield  {author} {,\ }%
  The thermodynamics of computation---a review,\ 
  \href{https://doi.org/10.1007/BF02084158}
  {\bibfield  {journal} {\bibinfo  {journal} {Int. J. Theor. Phys.}\ }\textbf {\bibinfo {volume} {21}},\ \bibinfo {pages} {905} (\bibinfo {year} {1982})}
  \BibitemShut {NoStop}
%
\bibitem [{\citenamefont {Lee}\ and\ \citenamefont {Peper}(2008)}]{lee2008brownian}
  \BibitemOpen
  \bibfield  {author} {
    \bibinfo {author} {\bibfnamefont {J.}~\bibnamefont  {Lee}}\ and\ 
    \bibinfo {author} {\bibfnamefont {F.}~\bibnamefont {Peper}},}
  \bibinfo {title} {On brownian cellular automata},\ in\ 
  \href{} {\emph  {\bibinfo {booktitle} {Automata}}}\   
  (\bibinfo {year} {2008})  \ pp.\ \bibinfo  {pages} {278--291}
  \BibitemShut {NoStop}
%
\bibitem [{\citenamefont {Peper}\ \emph {et~al.}(2013)
                \citenamefont {Peper},
                \citenamefont {Lee}, 
                \citenamefont {Carmona}, 
                \citenamefont {Cortadella},\  and\ 
                \citenamefont {Morita}}]{peper2013brownian}
  \BibitemOpen
  \bibinfo {author} {\bibfnamefont {F.}~\bibnamefont {Peper}}, 
  \bibinfo {author} {\bibfnamefont {J.}~\bibnamefont {Lee}}, 
  \bibinfo {author} {\bibfnamefont {J.}~\bibnamefont {Carmona}}, 
  \bibinfo {author} {\bibfnamefont {J.}~\bibnamefont {Cortadella}},\ and\ 
  \bibinfo {author} {\bibfnamefont {K.}~\bibnamefont {Morita}}\bibfield  {author} {,\ }%
  Brownian circuits: fundamentals,\ 
  \href{https://doi.org/10.1145/2422094.2422097} 
  {\bibfield  {journal} {\bibinfo  {journal} {ACM J. Emerg. Technol. Comput. Syst.}\ }\textbf  {\bibinfo {volume} {9}},\ \bibinfo {pages} {1} (\bibinfo {year}  {2013})}
  \BibitemShut {NoStop}
%
\bibitem [{\citenamefont {Lee}\ \emph {et~al.}(2016)
                \citenamefont {Lee},
                \citenamefont {Peper}, 
                \citenamefont {Cotofana}, 
                \citenamefont {Naruse},
                \citenamefont {Ohtsu}, 
                \citenamefont {Kawazoe}, 
                \citenamefont {Takahashi},
                \citenamefont {Shimokawa}, 
                \citenamefont {Kish},\ and\ 
                \citenamefont {Kubota}}]{lee2016brownian}
  \BibitemOpen
  \bibinfo {author} {\bibfnamefont {J.}~\bibnamefont {Lee}}, 
  \bibinfo {author} {\bibfnamefont {F.}~\bibnamefont {Peper}}, 
  \bibinfo {author} {\bibfnamefont {S.~D.}~\bibnamefont {Cotofana}}, 
  \bibinfo {author} {\bibfnamefont {M.}~\bibnamefont {Naruse}}, 
  \bibinfo {author} {\bibfnamefont {M.}~\bibnamefont {Ohtsu}}, 
  \bibinfo {author} {\bibfnamefont {T.}~\bibnamefont {Kawazoe}}, 
  \bibinfo {author} {\bibfnamefont {Y.}~\bibnamefont {Takahashi}}, 
  \bibinfo {author} {\bibfnamefont {T.}~\bibnamefont {Shimokawa}}, 
  \bibinfo {author} {\bibfnamefont {L.~B.}~\bibnamefont {Kish}},\ and\ 
  \bibinfo {author} {\bibfnamefont {T.}~\bibnamefont {Kubota}}\bibfield  {author} {,\ }%
  Brownian Circuits: Designs,\ 
  \href@noop {} {\bibfield  {journal} {\bibinfo  {journal}
  {Int. Journ. of Unconventional Computint}\ }\textbf {\bibinfo {volume} {12}} (\bibinfo {year} {2016})}
  \BibitemShut {NoStop}
%
\bibitem [{\citenamefont {Bennett}(1973)}]{bennett1973logical}
  \BibitemOpen
  \bibinfo {author} {\bibfnamefont {C.~H.}\ \bibnamefont {Bennett}}\bibfield {author} {,\ }%
  Logical reversibility of computation,\ 
  \href{https://doi.org/10.1147/rd.176.0525}
  {\bibfield  {journal} {\bibinfo  {journal} {IBM J. Res. Dev.}\ }\textbf {\bibinfo {volume} {17}},\ \bibinfo {pages} {525} (\bibinfo {year} {1973})}
  \BibitemShut {NoStop}
%
\bibitem [{\citenamefont {Feynman}(1986)}]{feynman1986quantum}
  \BibitemOpen
  \bibinfo {author} {\bibfnamefont {R.~P.}\ \bibnamefont {Feynman}}\bibfield{author} {,\ }%
  Quantum mechanical computers,\ 
  \href{https://doi.org/10.1515/9781400886975-036} 
  {\bibfield{journal} {\bibinfo  {journal} {Found. Phys.}\ }\textbf {\bibinfo {volume}{16}},\ \bibinfo {pages} {507} (\bibinfo {year} {1986})}
  \BibitemShut{NoStop}
%
\bibitem [{\citenamefont {Toffoli}(1980)}]{toffoli1980reversible}
  \BibitemOpen
  \bibfield  {author} {\bibinfo {author} {\bibfnamefont {T.}~\bibnamefont{Toffoli}},\ }%
  \bibinfo {title} {Reversible computing},\ in\ 
  \href{https://doi.org/10.1007/3-540-10003-2_104}
  {\emph {\bibinfo {booktitle} {International colloquium on automata, languages, and programming}}}\ 
  (\bibinfo {organization} {Springer},\ \bibinfo {year} {1980})\ pp.\ \bibinfo {pages} {632--644}
  \BibitemShut {NoStop}
%
\bibitem [{\citenamefont {Fredkin}\ and\ 
                \citenamefont {Toffoli}(1982)}]{fredkin1982conservative}
  \BibitemOpen
  \bibinfo {author} {\bibfnamefont {E.}~\bibnamefont {Fredkin}}\ and\ 
  \bibinfo {author} {\bibfnamefont {T.}~\bibnamefont {Toffoli}}\bibfield  {author} {,\ }%
  Conservative logic,\ 
  \href{https://doi.org/10.1007/BF01857727} 
  {\bibfield  {journal} {\bibinfo {journal} {Int. J. Theor. Phys.}\ }\textbf {\bibinfo  {volume} {21}},\ \bibinfo {pages} {219} (\bibinfo {year} {1982})}
  \BibitemShut{NoStop}
%
\bibitem [{\citenamefont {Landauer}(1961)}]{landauer1961irreversibility}
  \BibitemOpen
  \bibinfo {author} {\bibfnamefont {R.}~\bibnamefont {Landauer}}\bibfield  {author} {,\ }%
  Irreversibility and heat generation in the computing process,\ 
  \href{https://doi.org/10.1147/rd.53.0183}
   {\bibfield  {journal} {\bibinfo  {journal} {IBM J. Res. Dev.}\ }\textbf {\bibinfo {volume} {5}},\  \bibinfo {pages} {183} (\bibinfo {year} {1961})}
  \BibitemShut {NoStop}
%
\bibitem [{\citenamefont {Strasberg}\ \emph {et~al.}(2015)
                \citenamefont {Strasberg}, 
                \citenamefont {Cerrillo}, 
                \citenamefont {Schaller},\ and\
                \citenamefont {Brandes}}]{strasberg2015thermodynamics}
  \BibitemOpen
  \bibinfo {author} {\bibfnamefont {P.}~\bibnamefont {Strasberg}}, 
  \bibinfo {author} {\bibfnamefont {J.}~\bibnamefont {Cerrillo}}, 
  \bibinfo {author} {\bibfnamefont {G.}~\bibnamefont {Schaller}},\ and\ 
  \bibinfo {author} {\bibfnamefont {T.}~\bibnamefont {Brandes}}\bibfield  {author} {,\  }%
  Thermodynamics of stochastic Turing machines,\ 
  \href{https://doi.org/10.1103/PhysRevE.92.042104} 
  {\bibfield {journal} {\bibinfo  {journal} {Phys. Rev. E}\ }\textbf {\bibinfo {volume} {92}},\ \bibinfo {pages} {042104} (\bibinfo {year} {2015})}
  \BibitemShut {NoStop}
%
\bibitem [{\citenamefont {Norton}(2013)}]{norton2013brownian}
  \BibitemOpen
  \bibinfo {author} {\bibfnamefont {J.~D.}\ \bibnamefont {Norton}}\bibfield  {author} {,\ }%
  Brownian computation is thermodynamically irreversible,\
  \href{https://doi.org/10.1007/s10701-013-9753-1} 
  {\bibfield  {journal} {\bibinfo  {journal} {Found. Phys.}\ }\textbf {\bibinfo {volume} {43}},\ \bibinfo {pages} {1384}  (\bibinfo {year} {2013})}
  \BibitemShut {NoStop}
%
\bibitem [{\citenamefont {Ray}\ \emph {et~al.}(2021)
                \citenamefont {Ray},
                \citenamefont {Boyd}, 
                \citenamefont {Wimsatt},\ and\ 
                \citenamefont {Crutchfield}}]{ray2021non}
  \BibitemOpen
  \bibinfo {author} {\bibfnamefont {K.~J.}\ \bibnamefont {Ray}}, 
  \bibinfo {author} {\bibfnamefont {A.~B.}\ \bibnamefont {Boyd}}, 
  \bibinfo {author} {\bibfnamefont {G.~W.}\ \bibnamefont {Wimsatt}},\ and\ 
  \bibinfo {author} {\bibfnamefont {J.~P.}\ \bibnamefont {Crutchfield}}\bibfield  {author} {,\  }%
  Non-Markovian momentum computing: Thermodynamically efficient and computation universal,\ 
  \href{https://doi.org/10.1103/PhysRevResearch.3.023164} 
  {\bibfield  {journal} {\bibinfo  {journal} {Phys. Rev. Res.}\ }\textbf {\bibinfo {volume} {3}},\  \bibinfo {pages} {023164} (\bibinfo {year} {2021})}
  \BibitemShut {NoStop}
%
\bibitem [{\citenamefont {Owen}\ \emph {et~al.}(2019)
                \citenamefont {Owen},
                \citenamefont {Kolchinsky},\ and\ 
                \citenamefont {Wolpert}}]{owen2019number}
  \BibitemOpen
  \bibinfo {author} {\bibfnamefont {J.~A.}\ \bibnamefont {Owen}}, 
  \bibinfo {author} {\bibfnamefont {A.}~\bibnamefont {Kolchinsky}},\ and\ 
  \bibinfo {author} {\bibfnamefont {D.~H.}\ \bibnamefont {Wolpert}}\bibfield  {author} {,\ }%
  Number of hidden states needed to physically implement a given conditional distribution,\ 
  \href{https://doi.org/10.1088/1367-2630/aaf81d} 
  {\bibfield  {journal} {\bibinfo {journal} {New J. Phys.}\ }\textbf {\bibinfo {volume} {21}},\ \bibinfo {pages} {013022} (\bibinfo {year} {2019})}
  \BibitemShut {NoStop}
%
\bibitem [{\citenamefont {Stopnitzky}\ \emph {et~al.}(2019)
                \citenamefont {Stopnitzky}, 
                \citenamefont {Still}, 
                \citenamefont {Ouldridge},\ and\
                \citenamefont {Altenberg}}]{stopnitzky2019physical}
  \BibitemOpen
  \bibinfo {author} {\bibfnamefont {E.}~\bibnamefont {Stopnitzky}}, 
  \bibinfo {author} {\bibfnamefont {S.}~\bibnamefont {Still}}, 
  \bibinfo {author} {\bibfnamefont {T.~E.}\ \bibnamefont {Ouldridge}},\ and\ 
  \bibinfo {author} {\bibfnamefont {L.}~\bibnamefont {Altenberg}}\bibfield  {author} {,\  }
  Physical limitations of work extraction from temporal correlations,\ 
  \href{https://doi.org/10.1103/PhysRevE.99.042115} 
  {\bibfield  {journal} {\bibinfo {journal} {Phys. Rev. E}\ }\textbf {\bibinfo {volume} {99}},\ \bibinfo {pages} {042115} (\bibinfo {year} {2019})}
  \BibitemShut {NoStop}
%
\bibitem [{\citenamefont {Pal}\ \emph {et~al.}(2021)
                \citenamefont {Pal},
                \citenamefont {Reuveni},\ and\ 
                \citenamefont {Rahav}}]{pal2021thermodynamic}
  \BibitemOpen
  \bibinfo {author} {\bibfnamefont {A.}~\bibnamefont {Pal}}, 
  \bibinfo {author} {\bibfnamefont {S.}~\bibnamefont {Reuveni}},\ and\ 
  \bibinfo {author} {\bibfnamefont {S.}~\bibnamefont {Rahav}}\bibfield  {author} {,\  }%
  Thermodynamic uncertainty relation for first-passage times on Markov chains,\ 
  \href{https://doi.org/10.1103/PhysRevResearch.3.L032034} 
  {\bibfield  {journal} {\bibinfo  {journal} {Phys. Rev. Res.}\ }\textbf {\bibinfo {volume} {3}},\ \bibinfo {pages} {L032034} (\bibinfo {year} {2021})}
  \BibitemShut {NoStop}
%
\bibitem [{\citenamefont {Seifert}(2012)}]{seifert2012stochastic}
  \BibitemOpen
  \bibinfo {author} {\bibfnamefont {U.}~\bibnamefont {Seifert}}\bibfield  {author} {,\ }%
  Stochastic thermodynamics, fluctuation theorems and molecular machines,\ 
  \href{https://doi.org/10.1088/0034-4885/75/12/126001} 
  {\bibfield  {journal} {\bibinfo  {journal} {Rep. Prog. Phys.}\ }\textbf {\bibinfo {volume} {75}},\ \bibinfo {pages} {126001} (\bibinfo {year} {2012})}
  \BibitemShut {NoStop}
%
\bibitem [{\citenamefont {Utsumi}\ \emph {et~al.}(2022)
                \citenamefont {Utsumi},
                \citenamefont {Ito}, 
                \citenamefont {Golubev},\ and\ 
                \citenamefont {Peper}}]{utsumi2022computation}
  \BibitemOpen
  \bibfield  {author} {\bibinfo {author} {\bibfnamefont {Y.}~\bibnamefont {Utsumi}}, 
                               \bibinfo {author} {\bibfnamefont {Y.}~\bibnamefont {Ito}},
                               \bibinfo {author} {\bibfnamefont {D.}~\bibnamefont {Golubev}},\ and\ 
                               \bibinfo {author} {\bibfnamefont {F.}~\bibnamefont {Peper}},\ }%
  \bibinfo {title} {Computation time and thermodynamic uncertainty relation of Brownian circuits}, \
  \href{https://doi.org/10.48550/arXiv.2205.10735} 
  {arXiv:2205.10735}
  \BibitemShut {NoStop}
%
\bibitem [{\citenamefont {Utsumi}\ \emph {et~al.}(2023)
                \citenamefont {Utsumi},
                \citenamefont {Golubev},\ and\ 
                \citenamefont {Peper}}]{utsumi2023thermodynamic}
  \BibitemOpen
  \bibinfo {author} {\bibfnamefont {Y.}~\bibnamefont {Utsumi}}, 
  \bibinfo {author} {\bibfnamefont {D.}~\bibnamefont {Golubev}},\ and\ 
  \bibinfo {author} {\bibfnamefont {F.}~\bibnamefont {Peper}}\bibfield  {author} {,\ }%
  Thermodynamic cost of Brownian computers in the stochastic thermodynamics of resetting,\ 
  \href{https://doi.org/10.1140/epjs/s11734-023-00981-8} 
  {\bibfield  {journal} {\bibinfo  {journal} {Eur. Phys. J. Spec. Top.}\ }\textbf {\bibinfo {volume}{232}},\ \bibinfo {pages} {3259} (\bibinfo {year} {2023})}
  \BibitemShut {NoStop}
%
\bibitem [{\citenamefont {Risken}(1996)}]{risken1996fokker}
  \BibitemOpen
  \bibfield  {author} {\bibinfo {author} {\bibfnamefont {H.}~\bibnamefont {Risken}},\ }%
  \href{https://doi.org/10.1007/978-3-642-61544-3_6}
  {\bibinfo {title} {Fokker-planck equation for several variables; methods of solution}},\ in\  
  \emph {\bibinfo {booktitle} {The Fokker-Planck Equation: Methods of Solution and Applications}}\ 
  (\bibinfo {publisher} {Springer},\ \bibinfo {year} {1996})\ pp.\ \bibinfo {pages} {133--162}
  \BibitemShut {NoStop}
%
\bibitem [{\citenamefont {Magnasco}(1993)}]{magnasco1993forced}
  \BibitemOpen
  \bibinfo {author} {\bibfnamefont {M.~O.}\ \bibnamefont {Magnasco}}\bibfield {author} {,\ }%
  Forced thermal ratchets,\ 
  \href{https://doi.org/10.1103/PhysRevLett.71.1477} 
  {\bibfield  {journal} {\bibinfo {journal} {Phys. Rev. Lett.}\ }\textbf {\bibinfo {volume} {71}},\ \bibinfo {pages} {1477} (\bibinfo {year} {1993})}
  \BibitemShut {NoStop}
%
\bibitem [{\citenamefont {Astumian}\ and\ \citenamefont {Bier}(1994)}]{astumian1994}
  \BibitemOpen
  \bibinfo {author} {\bibfnamefont {R.~D.}\ \bibnamefont {Astumian}}\ and\
  \bibinfo {author} {\bibfnamefont {M.}~\bibnamefont {Bier}}\bibfield  {author} {,\ }%
  Fluctuation driven ratchets: Molecular motors,\ 
  \href{https://doi.org/10.1103/PhysRevLett.72.1766} 
  {\bibfield  {journal} {\bibinfo {journal} {Phys. Rev. Lett.}\ }\textbf {\bibinfo {volume} {72}},\ \bibinfo {pages} {1766} (\bibinfo {year} {1994})}
  \BibitemShut {NoStop}
%
\bibitem [{\citenamefont {Astumian}(1997)}]{astumian1997thermodynamics}
  \BibitemOpen
  \bibinfo {author} {\bibfnamefont {R.~D.}\ \bibnamefont {Astumian}}\bibfield {author} {,\ }%
  Thermodynamics and kinetics of a Brownian motor,\ 
  \href{https://doi.org/10.1126/science.276.5314.917} 
  {\bibfield  {journal} {\bibinfo  {journal} {Science}\ }\textbf {\bibinfo {volume} {276}},\ \bibinfo {pages} {917} (\bibinfo {year} {1997})}
  \BibitemShut {NoStop}
%
\bibitem [{\citenamefont {Reimann}(2002)}]{reimann2002brownian}
  \BibitemOpen
  \bibinfo {author} {\bibfnamefont {P.}~\bibnamefont {Reimann}}\bibfield {author} {,\ }%
  Brownian motors: noisy transport far from equilibrium,\
  \href{https://doi.org/10.1016/S0370-1573(01)00081-3} 
  {\bibfield  {journal} {\bibinfo  {journal} {Phys. Rep.}\ }\textbf {\bibinfo {volume} {361}},\ \bibinfo {pages} {57} (\bibinfo {year} {2002})}
  \BibitemShut {NoStop}
%
\bibitem [{\citenamefont {Nakade}\ \emph {et~al.}(2020)
                \citenamefont {Nakade},
                \citenamefont {Kanki}, 
                \citenamefont {Tanaka},\ and\ 
                \citenamefont {Petrosky}}]{nakade2020anomalous}
  \BibitemOpen
  \bibinfo {author} {\bibfnamefont {S.}~\bibnamefont {Nakade}}, 
  \bibinfo {author} {\bibfnamefont {K.}~\bibnamefont {Kanki}}, 
  \bibinfo {author} {\bibfnamefont {S.}~\bibnamefont {Tanaka}},\ and\ 
  \bibinfo {author} {\bibfnamefont {T.}~\bibnamefont {Petrosky}}\bibfield  {author} {,\ }%
  Anomalous diffusion of a quantum Brownian particle in a one-dimensional molecular chain,\
  \href{https://doi.org/10.1103/PhysRevE.102.032137} 
  {\bibfield  {journal} {\bibinfo  {journal} {Phys. Rev. E}\ }\textbf {\bibinfo {volume} {102}},\ \bibinfo {pages} {032137} (\bibinfo {year} {2020})}
  \BibitemShut {NoStop}
%
\bibitem [{\citenamefont {Feynman}\ \emph {et~al.}(1963)
                \citenamefont {Feynman}, 
                \citenamefont {Leighton},\ and\ 
                \citenamefont {Sands}}]{feynman1963the}%
  \BibitemOpen
  \bibfield  {author} {\bibinfo {author} {\bibfnamefont {R.~P.}\ \bibnamefont {Feynman}}, 
  \bibinfo {author} {\bibfnamefont {R.~B.}\ \bibnamefont {Leighton}},\ and\ 
  \bibinfo {author} {\bibfnamefont {M.}~\bibnamefont {Sands}},\ }
  \href@noop {} 
  {\emph {\bibinfo {title} {The Feynman Lectures on Physics}}},\ 
  Vol.~\bibinfo {volume} {1}\ (\bibinfo  {publisher} {Addison-Wesley, Reading, MA},\ \bibinfo {year} {1963})
  \BibitemShut {NoStop}
%
\bibitem [{\citenamefont {Szilard}(1929)}]{Szilard:1929aa}
  \BibitemOpen
  \bibinfo {author} {\bibfnamefont {L.}~\bibnamefont {Szilard}}\bibfield {author} {,\ }%
  {\"U}ber die Entropieverminderung in einem thermodynamischen System bei Eingriffen intelligenter Wesen,\ 
  \href{https://doi.org/10.1007/BF01341281} 
  {\bibfield  {journal} {\bibinfo {journal} {Z. Phys.}\ }\textbf {\bibinfo {volume} {53}},\ \bibinfo {pages} {840} (\bibinfo {year} {1929})};\
  English translation: On the decrease of entropy in a thermodynamic system by the intervention of intelligent beings, 
  \href{https://doi.org/10.1002/bs.3830090402}
  {Behavioral Science, \textbf{9}, 301 (1964)}
\BibitemShut {NoStop}
%
\bibitem [{\citenamefont {Sagawa}(2018)}]{sagawa2018second}
  \BibitemOpen
  \bibfield  {author} {\bibinfo {author} {\bibfnamefont {T.}~\bibnamefont {Sagawa}},\ }%
  \href{https://doi.org/10.1007/978-3-319-93458-7_3}
  {\bibinfo {title} {Second law, entropy production, and reversibility in thermodynamics of information}},\ in\  
  \emph {\bibinfo {booktitle} {Energy Limits in Computation: A Review of Landauer's Principle, Theory and Experiments}}\ 
  (\bibinfo  {publisher} {Springer},\ \bibinfo {year} {2018})\ pp.\ \bibinfo {pages} {101--139}
  \BibitemShut {NoStop}
%
\bibitem [{\citenamefont {Nakade}\ \emph {et~al.}(2021)
                \citenamefont {Nakade},
                \citenamefont {Kanki}, 
                \citenamefont {Tanaka},\ and\ 
                \citenamefont {Petrosky}}]{nakade2021anomalous}
  \BibitemOpen
  \bibinfo {author} {\bibfnamefont {S.}~\bibnamefont {Nakade}}, 
  \bibinfo {author} {\bibfnamefont {K.}~\bibnamefont {Kanki}}, 
  \bibinfo {author} {\bibfnamefont {S.}~\bibnamefont {Tanaka}},\ and\ 
  \bibinfo {author} {\bibfnamefont {T.}~\bibnamefont {Petrosky}}\bibfield  {author} {,\ }%
  Anomalous Diffusion with an Apparently Negative Diffusion Coefficient in a One-Dimensional Quantum Molecular Chain Model,\ 
  \href{https://doi.org/10.3390/sym13030506} 
  {\bibfield  {journal} {\bibinfo {journal} {Symmetry}\ }\textbf {\bibinfo {volume} {13}},\ \bibinfo {pages} {506} (\bibinfo {year} {2021})}
  \BibitemShut {NoStop}
%
\bibitem [{\citenamefont {Tanaka}\ \emph {et~al.}(2009)
                \citenamefont {Tanaka},
                \citenamefont {Kanki},\ and\ 
                \citenamefont {Petrosky}}]{tanaka2009emergence}
  \BibitemOpen
  \bibinfo {author} {\bibfnamefont {S.}~\bibnamefont {Tanaka}}, 
  \bibinfo {author} {\bibfnamefont {K.}~\bibnamefont {Kanki}},\ and\ 
  \bibinfo {author} {\bibfnamefont {T.}~\bibnamefont {Petrosky}}\bibfield  {author} {,\  }%
  Emergence of quantum hydrodynamic sound mode of a quantum Brownian particle in a one-dimensional molecular chain,\ 
  \href{https://doi.org/10.1103/PhysRevB.80.094304} 
  {\bibfield  {journal} {\bibinfo  {journal} {Phys. Rev. B}\ }\textbf {\bibinfo {volume} {80}},\ \bibinfo {pages} {094304} (\bibinfo {year} {2009})}
  \BibitemShut {NoStop}
%
\bibitem [{\citenamefont {Petrosky}\ \emph {et~al.}(2010)
                \citenamefont {Petrosky}, 
                \citenamefont {Hatano}, 
                \citenamefont {Kanki},\ and\ 
                \citenamefont {Tanaka}}]{petrosky2010hofstadter}
  \BibitemOpen
  \bibinfo {author} {\bibfnamefont {T.}~\bibnamefont {Petrosky}}, 
  \bibinfo {author} {\bibfnamefont {N.}~\bibnamefont {Hatano}}, 
  \bibinfo {author} {\bibfnamefont {K.}~\bibnamefont {Kanki}},\ and\ 
  \bibinfo {author} {\bibfnamefont {S.}~\bibnamefont {Tanaka}}\bibfield  {author} {,\  }%
  Hofstadter's Butterfly Type of Singular Spectrum of a Collision Operator for a Model of Molecular Chains,\ 
  \href{https://doi.org/10.1143/PTPS.184.457} 
  {\bibfield  {journal} {\bibinfo {journal} {Prog. Theor. Phys. Suppl.}\ }\textbf {\bibinfo {volume} {184}},\ \bibinfo {pages} {457} (\bibinfo {year} {2010})}
  \BibitemShut {NoStop}
%
\bibitem [{\citenamefont {R{\'e}sibois}\ and\ 
                \citenamefont {de~Leener}(1977)}]{1977PResiboisMdeLeenery}
  \BibitemOpen
  \bibfield  {author} {\bibinfo {author} {\bibfnamefont {P.}~\bibnamefont {R{\'e}sibois}}\ and\ 
  \bibinfo {author} {\bibfnamefont {M.}~\bibnamefont {de~Leener}},\ }%
  \href@noop {} 
  {\emph {\bibinfo {title} {Classical Kinetic Theory of Fluids}}}\ 
  (\bibinfo  {publisher} {John Wiley \& Sons},\ \bibinfo {address} {New York},\ \bibinfo {year} {1977})
  \BibitemShut {NoStop}
%
\bibitem [{\citenamefont {Cover}(1999)}]{cover1999elements}
  \BibitemOpen
  \bibfield  {author} {\bibinfo {author} {\bibfnamefont {T.~M.}\ \bibnamefont {Cover}},\ }%
  \href@noop {} 
  {\emph {\bibinfo {title} {Elements of information theory}}}\ (\bibinfo  {publisher} {John Wiley \& Sons},\ \bibinfo {year} {1999})
  \BibitemShut {NoStop}
%
\bibitem [{\citenamefont {Tay}\ \emph {et~al.}(2011)
                \citenamefont {Tay},
                \citenamefont {Kanki}, 
                \citenamefont {Tanaka},\ and\ 
                \citenamefont {Petrosky}}]{tay2011band}
  \BibitemOpen
  \bibinfo {author} {\bibfnamefont {B.}~\bibnamefont {Tay}}, 
  \bibinfo {author} {\bibfnamefont {K.}~\bibnamefont {Kanki}}, 
  \bibinfo {author} {\bibfnamefont {S.}~\bibnamefont {Tanaka}},\ and\ 
  \bibinfo {author} {\bibfnamefont {T.}~\bibnamefont {Petrosky}}\bibfield  {author} {,\ }%
  Band structure and accumulation point in the spectrum of quantum collision operator in a one-dimensional molecular chain,\ 
  \href{https://doi.org/10.1063/1.3553201} 
  {\bibfield  {journal} {\bibinfo  {journal} {J. Math. Phys.}\ }\textbf {\bibinfo {volume} {52}} (\bibinfo {year} {2011})}
  \BibitemShut {NoStop}
%
\bibitem [{\citenamefont {Prigogine}(1962)}]{prigogine1962non}
  \BibitemOpen
  \bibfield  {author} {\bibinfo {author} {\bibfnamefont {I.}~\bibnamefont {Prigogine}},\ }%
  \href@noop {} 
  {\emph {\bibinfo {title} {Nonequilibrium statistical mechanics}}}\ 
  (\bibinfo  {publisher} {John Willey \& Sons},\ \bibinfo {year} {1962})
  \BibitemShut {NoStop}
%
\bibitem [{\citenamefont {Davydov}(1982)}]{davydov1982solitons}
  \BibitemOpen
  \bibinfo {author} {\bibfnamefont {A.~S.}\ \bibnamefont {Davydov}}\bibfield {author} {,\ }%
  Solitons in quasi-one-dimensional molecular structures,\
  \href{https://doi.org/10.1070/PU1982v025n12ABEH005012} 
  {\bibfield  {journal} {\bibinfo  {journal} {Sov. Phys.--Uspeki}\ }\textbf {\bibinfo {volume} {25}},\ \bibinfo {pages} {898} (\bibinfo {year} {1982})}
  \BibitemShut {NoStop}
%
\bibitem [{\citenamefont {Scott}(1992)}]{scott1992davydov}
  \BibitemOpen
  \bibinfo {author} {\bibfnamefont {A.}~\bibnamefont {Scott}}\bibfield  {author} {,\ }%
  Davydov's soliton,\ 
  \href{https://doi.org/10.1016/0370-1573(92)90093-F} 
  {\bibfield  {journal} {\bibinfo {journal} {Phys. Rep.}\ }\textbf {\bibinfo {volume} {217}},\ \bibinfo {pages} {1} (\bibinfo {year} {1992})}
  \BibitemShut {NoStop}
%
\bibitem [{\citenamefont {Mahan}(2013)}]{mahan1993many}
  \BibitemOpen
  \bibfield  {author} {\bibinfo {author} {\bibfnamefont {G.~D.}\ \bibnamefont {Mahan}},\ }%
  \href@noop {} 
  {\emph {\bibinfo {title} {Many-particle physics}}},\ 
  \bibinfo {edition} {2nd}\ ed.\ (\bibinfo  {publisher} {Plenum Press},\ \bibinfo {address} {New York and London},\ \bibinfo {year} {2013})
  \BibitemShut {NoStop}
%
\bibitem [{\citenamefont {Petrosky}\ and\ 
                \citenamefont {Prigogine}(1997)}]{petrosky1997liouville}
  \BibitemOpen
  \bibinfo {author} {\bibfnamefont {T.}~\bibnamefont {Petrosky}}\ and\ 
  \bibinfo {author} {\bibfnamefont {I.}~\bibnamefont {Prigogine}}\bibfield  {author} {,\ }%
  The Liouville space extension of quantum mechanics,\ 
  \href{https://doi.org/10.1002/9780470141588}
  {\bibfield  {journal} {\bibinfo  {journal} {Adv. Chem. Phys.}\ }\textbf {\bibinfo {volume} {99}},\ \bibinfo {pages} {1} (\bibinfo {year} {1997})}
  \BibitemShut {NoStop}
%
\bibitem [{\citenamefont {Petrosky}\ and\ 
                \citenamefont {Prigogine}(1996)}]{petrosky1996poincare}
  \BibitemOpen
  \bibinfo {author} {\bibfnamefont {T.}~\bibnamefont {Petrosky}}\ and\ 
  \bibinfo {author} {\bibfnamefont {I.}~\bibnamefont {Prigogine}}\bibfield  {author} {,\ }%
  Poincar{\'e} resonances and the extension of classical dynamics,\
  \href{https://doi.org/10.1016/0960-0779(95)00042-9} 
  {\bibfield  {journal} {\bibinfo  {journal} {Chaos, Solitons \& Fractals}\ }\textbf {\bibinfo {volume} {7}},\ \bibinfo {pages} {441} (\bibinfo {year} {1996})}
  \BibitemShut {NoStop}
%
\bibitem [{\citenamefont {Hudson}(1974)}]{hudson1974wigner}
  \BibitemOpen
  \bibinfo {author} {\bibfnamefont {R.~L.}\ \bibnamefont {Hudson}}\bibfield {author} {,\ }%
  When is the Wigner quasi-probability density non-negative?,\
  \href{https://doi.org/10.1016/0034-4877(74)90007-X} 
  {\bibfield  {journal} {\bibinfo  {journal} {Rep. Math. Phys.}\ }\textbf {\bibinfo {volume} {6}},\ \bibinfo {pages} {249} (\bibinfo {year} {1974})}
  \BibitemShut {NoStop}
%
\bibitem [{\citenamefont {Leonhardt}\ and\ 
                \citenamefont {Paul}(1995)}]{leonhardt1995measuring}
  \BibitemOpen
  \bibinfo {author} {\bibfnamefont {U.}~\bibnamefont {Leonhardt}}\ and\ 
  \bibinfo {author} {\bibfnamefont {H.}~\bibnamefont {Paul}}\bibfield  {author} {,\ }%
  Measuring the quantum state of light,\ 
  \href{https://doi.org/10.1016/0079-6727(94)00007-L} 
  {\bibfield  {journal} {\bibinfo  {journal} {Prog. Quantum Electron.}\ }\textbf {\bibinfo {volume} {19}},\ \bibinfo {pages} {89} (\bibinfo {year} {1995})}
  \BibitemShut {NoStop}
%
\bibitem [{\citenamefont {Shannon}(1948)}]{shannon1948mathematical}
  \BibitemOpen
  \bibinfo {author} {\bibfnamefont {C.~E.}\ \bibnamefont {Shannon}}\bibfield {author} {,\ }%
  A mathematical theory of communication,\ 
  \href{https://doi.org/10.1145/584091.584093}
  {\bibfield  {journal} {\bibinfo  {journal} {Bell Syst. Tech. J.}\ }\textbf {\bibinfo {volume} {27}},\ \bibinfo {pages} {379} (\bibinfo {year} {1948})}
  \BibitemShut {NoStop}
%
\bibitem [{\citenamefont {Boyd}\ \emph {et~al.}(2016)
                \citenamefont {Boyd},  
                \citenamefont {Mandal},\ and\ 
                \citenamefont {Crutchfield}}]{boyd2016identifying}
  \BibitemOpen
  \bibinfo {author} {\bibfnamefont {A.~B.}\ \bibnamefont {Boyd}}, 
  \bibinfo {author} {\bibfnamefont {D.}~\bibnamefont {Mandal}},\ and\ 
  \bibinfo {author} {\bibfnamefont {J.~P.}\ \bibnamefont {Crutchfield}}\bibfield  {author} {,\ }%
  Identifying functional thermodynamics in autonomous Maxwellian ratchets,\
  \href{https://doi.org/10.1088/1367-2630/18/2/023049} 
  {\bibfield  {journal} {\bibinfo  {journal} {New J. Phys.}\ }\textbf {\bibinfo {volume} {18}},\ \bibinfo {pages} {023049} (\bibinfo {year} {2016})}
  \BibitemShut {NoStop}
%
\bibitem [{\citenamefont {S{\'a}nchez}\ \emph {et~al.}(2019)
                \citenamefont {S{\'a}nchez}, 
                \citenamefont {Samuelsson},\ and\ 
                \citenamefont {Potts}}]{sanchez2019autonomous}
  \BibitemOpen
  \bibinfo {author} {\bibfnamefont {R.}~\bibnamefont {S{\'a}nchez}}, 
  \bibinfo {author} {\bibfnamefont {P.}~\bibnamefont {Samuelsson}},\ and\ 
  \bibinfo {author} {\bibfnamefont {P.~P.}\ \bibnamefont {Potts}}\bibfield  {author} {,\ }%
  Autonomous conversion of information to work in quantum dots,\ 
  \href{https://doi.org/10.1103/PhysRevResearch.1.033066} 
  {\bibfield  {journal} {\bibinfo  {journal} {Phys. Rev. Res.}\ }\textbf {\bibinfo {volume} {1}},\ \bibinfo {pages} {033066} (\bibinfo {year} {2019})}
  \BibitemShut {NoStop}
%
\bibitem [{\citenamefont {Christiansen}\ and\ 
                \citenamefont {Scott}(1990)}]{christiansen1990davydov}
  \BibitemOpen
  \bibfield  {author} {\bibinfo {author} {\bibfnamefont {P.~L.}\ \bibnamefont {Christiansen}}\ and\ 
  \bibinfo {author} {\bibfnamefont {A.~C.}\ \bibnamefont {Scott}},\ }
  \href@noop {} 
  {\emph {\bibinfo {title} {Davydov's soliton revisited: self-trapping of vibrational energy in protein}}}\ 
  (\bibinfo {publisher} {Plenum, New York},\ \bibinfo {year} {1990})
  \BibitemShut {NoStop}
%
\bibitem [{\citenamefont {Ray}\ and\ 
                \citenamefont {Crutchfield}(2023)}]{ray2023gigahertz}
  \BibitemOpen
  \bibinfo {author} {\bibfnamefont {K.~J.}\ \bibnamefont {Ray}}\ and\ 
  \bibinfo {author} {\bibfnamefont {J.~P.}\ \bibnamefont {Crutchfield}}\bibfield {author} {,\ }%
  Gigahertz sub-Landauer momentum computing,\ 
  \href{https://doi.org/10.1103/PhysRevApplied.19.014049}
  {\bibfield  {journal} {\bibinfo  {journal} {Phys. Rev. Appl.}\ }\textbf {\bibinfo {volume} {19}},\ \bibinfo {pages} {014049} (\bibinfo {year} {2023})}
  \BibitemShut {NoStop}
%
\bibitem [{\citenamefont {Calcaterra}\ and\ 
                \citenamefont {Boldt}(2008)}]{calcaterra2008approximating}
  \BibitemOpen
  \bibfield  {author} {\bibinfo {author} {\bibfnamefont {C.}~\bibnamefont {Calcaterra}}\ and\ 
  \bibinfo {author} {\bibfnamefont {A.}~\bibnamefont {Boldt}},\ }
  \bibinfo {title} {Approximating with gaussians},\
  \href{https://doi.org/10.48550/arXiv.0805.3795}
  {arXiv:0805.3795}
  \BibitemShut {NoStop}
\end{thebibliography}

%

\end{document}